\documentclass[tbtags]{JHEP3}

\usepackage{graphicx}           % zum Einbinden von Grafiken
\usepackage{psfrag}             % zur Ersetzung von Texten in Grafiken
\usepackage{amsmath}            % liefert viele Tools für Math-Mode

\DeclareMathOperator{\tr}{tr}                               % trace operator
\newcommand{\be}{\begin{equation}}
\newcommand{\ee}{\end{equation}}
\newcommand{\ba}{\begin{eqnarray}}
\newcommand{\ea}{\end{eqnarray}}
\newcommand{\gym}{g_{\scriptscriptstyle\mathrm{YM}}}        % coupling constant g_YM
\newcommand{\N}{\mathcal{N}}                                % number of susy's
\newcommand{\Z}{\mathbb{Z}}                                 % set of integers
\newcommand{\R}{\mathbb{R}}                                 % set of real numbers
\newcommand{\eps}{\varepsilon}                              % shorthand \eps for \varepsilon
\newcommand{\order}[1]{\mathcal{O}\left(#1\right)}          % order O(..)
\newcommand{\comm}[2]{[#1,#2]}                              % commutator
\newcommand{\grU}{\mathrm{U}}                               % group U
\newcommand{\grSU}{\mathrm{SU}}                             % group SU
\newcommand{\grSO}{\mathrm{SO}}                             % group SO
\newcommand{\vev}[1]{\bigl<#1\bigr>}                        % Vacuum expectation value
\newcommand{\ket}[1]{\bigl|#1\bigr>}                        % Ket vector
\newcommand{\D}{\mathcal{D}}                                % Operator D
\renewcommand{\O}{\mathcal{O}}                              % Operator O

\newcommand{\vertex}[7]{\psfrag{1}{#2} \psfrag{2}{#3}\psfrag{3}{#4}\psfrag{4}{#5}\psfrag{a}{#6}\psfrag{b}{#7}
                        \raisebox{-6mm}{\includegraphics*[scale=.6]{#1.eps}}}
\newcommand{\h}[1]{\hphantom{#1}}    % for placement of objects in figures

\title{Conformal Dimensions of Two-Derivative BMN Operators}
\author{Thomas Klose \\
        Max-Planck-Institut f\"ur Gravitationsphysik \\
        Albert-Einstein-Institut\\
        Am M\"uhlenberg 1, D-14476 Golm, Germany \\
        E-mail: \email{thomas.klose@aei.mpg.de}}
\abstract{We compute the anomalous dimensions of BMN operators with two covariant derivative impurities at the planar %%@
level up to first order in the effective coupling $\lambda'$. The result equals those for two scalar impurities as well %%@
as for mixed scalar and vector impurities given in the literature. Though the results are the same, the computation is %%@
very different from the scalar case. This is basically due to the existence of a non-vanishing overlap between the %%@
derivative impurity and the ``background'' field $Z$. We present details of these differences and their consequences.}
\preprint{hep-th/0301150 \\
          AEI-2003-007}

\begin{document}

%%%%%%%%%%%%%%%%%%%%%%%%%%%%
%%%%%%% Introduction %%%%%%%
%%%%%%%%%%%%%%%%%%%%%%%%%%%%

\section{Introduction}

The Berenstein-Maldacena-Nastase (BMN) correspondence \cite{Berenstein:2002jq} is a limit of the AdS/CFT duality %%@
\cite{Aharony:1999ti} between type IIB superstring theory on $AdS_5 \times S^5$ and $\N = 4$ supersymmetric Yang-Mills %%@
theory on $\R^4$. The novel feature is that there exists a regime of the effective coupling strengths where both %%@
theories can be treated perturbatively at the same time. On the string theory side the BMN limit amounts to a Penrose %%@
limit \cite{Penrose} of $AdS_5 \times S^5$ and leads to a plane-wave geometry \cite{Blau:2002dy} which is labeled by a %%@
parameter $\mu$ of unit mass dimension. It is the geometry which a particle experiences that travels with large angular %%@
momentum $J$ along a circle $S^1$ on the 5-sphere $S^5$. Choosing one particular circle corresponds in the super %%@
Yang-Mills theory to singling out a $\grU(1)$ subgroup of the $\grSU(4)$ $R$-symmetry group. In consequence, the string %%@
states are represented by operators with large $U(1)_R$ charge $J$ which are commonly named BMN operators.

The correspondence is made more precise by the following dictionary. The string theory Hamiltonian $\hat %%@
H^{\mbox{\scriptsize l.c.}}$ in light-cone gauge is identified with the Yang-Mills dilatation operator $\hat D$ and the %%@
generator $\hat J$ of $\grU(1)_R$ transformations by the relation 
\be \label{eqn:BMN_correspondence}
\frac{1}{\mu} \hat H^{\mbox{\scriptsize l.c.}} \stackrel{!}{=} \hat D - \hat J \; .
\ee
In other words, this is an identification of the string energy with the conformal dimension minus the $R$-charge of the %%@
BMN operator. The parameters of both theories are related in the following way
\be \label{eqn:parameters}
\lambda' := \frac{\gym^2 N}{J^2} = \frac{1}{(\mu p^+ \alpha')^2} \qquad , \qquad g_2 := \frac{J^2}{N} = 4\pi g_s (\mu %%@
p^+ \alpha')^2 \; , 
\ee
where $\gym$ is the Yang-Mills coupling constant, $N$ is the rank of the gauge group $\grU(N)$, $g_s$ is string %%@
coupling constant and $p^+$ the string light-cone momentum. In the BMN limit one takes
\be \label{eqn:BMN_limit}
N,J \to \infty \quad\mbox{such that $\lambda',g_2$ fixed} \; .
\ee
Hence $\lambda'$ and $g_2$ are the relevant parameters in the BMN limit. They control the quantum loop and the genus %%@
expansion, respectively. As opposed to the 't Hooft limit \cite{'tHooft:1973jz} ($N\to\infty$ with %%@
$\lambda:=\gym^2N=\mbox{fixed}$) it has been found \cite{Kristjansen:2002bb}\cite{Constable:2002hw} that in the BMN %%@
limit also non-planar diagrams contribute on the gauge theory side. At each order in $\lambda'$ there is a series of %%@
diagrams with increasing genera controlled by $g_2$. In order to simplify computations one could concentrate on planar %%@
diagrams by setting $g_2=0$. On the string theory side the two expansions have the following significance. The %%@
light-cone string in the plane-wave background is in fact a massive string and its mass is given by $\mu \sim %%@
1/\sqrt{\lambda'}$. The other parameter $g_2 \sim g_s$ determines the strength of string splitting and joining. \\

The relationship \eqref{eqn:BMN_correspondence} has been written as an operator equation. However, both sides act on %%@
very different Hilbert spaces. Therefore it is necessary to know how string states are translated into BMN operators %%@
and vice versa. This question is a subject of current research \cite{Gross:2002mh}--\cite{Gomis:2003kj}. Since the %%@
original proposal of BMN a lot of work has been done to understand the correspondence in more detail especially at the %%@
interacting level. Some further references for investigations on the gauge theory side are %%@
\cite{Gross:2002su}--\cite{Beisert:2002ff} and for the string theory side see e.~g. %%@
\cite{Metsaev:2001bj}--\cite{Chu:2003qd}. In \cite{Verlinde:2002ig}\cite{Zhou:2002mi}\cite{Vaman:2002ka} a string bit %%@
formalism has been developed which interpolates between the two sides.

In the following we summarize the mapping between string states and BMN operators at the planar level ($g_2=0$) where %%@
the dictionary is well established. In this case the Hamiltonian simplifies to that of a non-interacting massive string %%@
which can be solved exactly \cite{Metsaev:2001bj}\cite{Metsaev:2002re}. In terms of occupation numbers $N_k$ the %%@
Hamiltonian reads
\be \label{eqn:hamiltonian_planar}
\frac{1}{\mu} \hat H^{\mbox{\scriptsize l.c.}} = \sum_{k=-\infty}^\infty N_k \sqrt{1 + \lambda' k^2} \; ,
\ee
where $N_k$ counts all bosonic as well as all fermionic\footnote{In the following we will mainly concentrate on the %%@
bosonic part of the theory.} excitations of mode number $k$. As opposed to the ordinary case, for the massive string %%@
there is a pair of (bosonic) creation and annihilation operators $a_k^{\dag i}, a_k^i$ independent for all modes %%@
$-\infty<k<\infty$. The index $i$ runs from 1 to 8 corresponding to the number of independent oscillators in light-cone %%@
gauge. 

In the diagonal basis underlying \eqref{eqn:hamiltonian_planar} the relation \eqref{eqn:BMN_correspondence} becomes
\be \label{eqn:BMN_correspondence_planar}
\sum_{k=-\infty}^\infty N_k \sqrt{1 + \lambda' k^2} = \Delta - J \; ,
\ee
where $\Delta = \Delta^{(0)} + \delta\Delta$  is the conformal dimension of the corresponding operator. It is given by %%@
the engineering conformal dimension $\Delta^{(0)}$ corrected by the anomalous dimension $\delta\Delta$ which is due to %%@
quantum effects. 

The (single) string vacuum $\ket{0,p^+}$ has zero energy and therefore should be represented by an operator whose %%@
conformal dimension equals its $R$-charge. The operator with this property, proposed by BMN \cite{Berenstein:2002jq}, %%@
is
\be \label{eqn:BMN_vacuum_operator}
\O(x) \sim \tr Z^J(x) \; ,
\ee
where $Z$ is that combination of Yang-Mills scalar fields with unit $R$-charge, cf. appendix %%@
\ref{sec:appendix_yang_mills}. This operator $\O$ is protected, i.~e. its engineering conformal dimension does not %%@
receive quantum corrections. Table \ref{tab:engineering_conformal_dimensions_and_r_charges} summarizes the engineering %%@
conformal dimensions and $R$-charges of the fields of the Yang-Mills theory. For composite operators these values add %%@
up. 
\TABLE{
\begin{tabular}{l|c|c|c|c|c|c}
               & \h{-1/2}       & \h{-1/2} & \h{-1/2}& \h{-1/2}      & \h{-1/2} \\[-10mm]
               & $Z$ & $\bar Z$ & $\phi_i$ & $D_\mu$ & $\psi_\alpha$ & $\bar\psi_\alpha$ \\ \hline
$\Delta^{(0)}$ & 1   &  1       &   1      &   1     &  3/2          &   3/2 \vphantom{\rule[1mm]{1mm}{3mm}} \\ 
$J$            & 1   & -1       &   0      &   0     &  1/2          &  -1/2
\end{tabular}
\caption{Engineering conformal dimensions $\Delta^{(0)}$ and $R$-charges $J$}
\label{tab:engineering_conformal_dimensions_and_r_charges}
}

The excited states
\be
a_{k_1}^{\dag i_1} \cdots a_{k_l}^{\dag i_l} \ket{0,p^+}
\ee 
are divided into supergravity states where only zero mode oscillators are applied $(k_1=\ldots=k_l=0)$ and %%@
non-supergravity states where the level matching condition demands $k_1+\ldots+k_l=0$. BMN operators corresponding to %%@
excited states are derived from \eqref{eqn:BMN_vacuum_operator} by inserting so-called ``impurities'' in-between the %%@
$Z$'s. Excitations along the directions $i=5,6,7,8$ (originally on $S^5$) are generated by scalar field impurities %%@
$\phi_{i-4}$ whereas the insertion of covariant derivatives $D_\mu$ models excitations along the remaining directions %%@
$\mu=1,2,3,4$ (originally on $AdS_5$). Fermionic excitations are obtained by inserting gaugino fields $\psi_\alpha$ %%@
$(\alpha = 1,\ldots,8)$. For higher excitations where there is more than one impurity one sums over all possible %%@
insertion points. If one performs this sum with appropriate phase factors which encode the mode numbers, one can %%@
realize non-supergravity modes. Some examples are given in table \ref{tab:correspondence_examples}. 
\TABLE{
\begin{tabular}{lllll}
String state & BMN operator & occupation no & (bare) energy \\[1mm] \hline
$\ket{0,p^+}$ & $\tr Z^J$ & $N_k = 0$ & $\Delta - J = 0$ \vphantom{\rule[1mm]{1mm}{3mm}} \\[2mm]
$a_{0}^{\dag i} \ket{0,p^+}$ & $\tr \phi_{i-4} Z^J \; , \quad (i=5,\ldots,8)$ & $N_k = \delta_{k0}$ & $\Delta - J = 1$ %%@
\\[2mm]
$a_{0}^{\dag \mu} \ket{0,p^+}$ & $\tr (D_\mu Z) Z^{J-1} \; , \quad (\mu=1,\ldots,4)$ & $N_k = \delta_{k0}$ & $\Delta - %%@
J = 1$ \\[2mm]
$a_{0}^{\dag i} a_{0}^{\dag \mu} \ket{0,p^+}$ &
$\sum_{p=0}^{J-1} \tr \phi_{i-4} Z^p (D_\mu Z) Z^{J-p-1}$ & $N_k = 2\delta_{k0}$ & $\Delta - J = 2$ \\
          & $\quad + \tr (D_\mu \phi_{i-4}) Z^J$ \\[2mm]
$a_{n}^{\dag 5} a_{-n}^{\dag 6} \ket{0,p^+}$ &
$\sum_{p=0}^J \tr \phi_1 Z^p \phi_2 Z^{J-p} \: e^{2\pi i n p/J}$ & $N_k = \delta_{kn}+\delta_{k,-n}$ & $\Delta^{(0)} - %%@
J = 2$
\end{tabular}
\caption{Examples of string operator correspondence. All states have light-cone momentum $p^+$ corresponding to %%@
$R$-charge $J$.}
\label{tab:correspondence_examples}
}

Eq. \eqref{eqn:BMN_correspondence_planar} has been verified for operators with scalar field impurities to second order %%@
in $\lambda'$ in \cite{Gross:2002su} and to all orders in \cite{Santambrogio:2002sb}, and for mixed scalar field and %%@
covariant derivative impurities up to $\order{\lambda'}$ in \cite{Gursoy:2002yy}. In this paper we confirm %%@
\eqref{eqn:BMN_correspondence_planar} to first order in $\lambda'$ for operators with two covariant derivative %%@
impurities. On the string theory side (at least at the non-interacting level) there is no difference between the %%@
oscillation modes in the directions $\mu=1,2,3,4$ and $i=5,6,7,8$. However the gauge theory calculations with %%@
derivative impurities are very different from the case of scalar impurities, in fact they are much more complex. This %%@
is ultimately due to the overlap between derivative impurity and ``background'' field $Z$ which vanishes for scalar %%@
impurities. Furthermore one has to cope with more diagrams stemming from insertions of the gauge field that is %%@
contained in the derivative impurity. Therefore it is very interesting to see how the result emerges in this more %%@
intricate case.

The paper is organized as follows. In the next section we introduce the operators we want to study and discuss some of %%@
their properties. In section \ref{sec:one_loop_computation} we present in detail the computation of their one-loop %%@
anomalous dimensions in the planar BMN limit. We conclude by summarizing the specialties of this computation in section %%@
\ref{sec:conclusion}. Our notation and conventions can be found in the appendix.

%%%%%%%%%%%%%%%%%%%%%%%%%%%%%%%%%%%%%%%%%%%%%%%%%%%%%%%%%%%%%%%%%%%%%%%%%%%
%%%%%%% Two-derivative BMN operators and their conformal dimensions %%%%%%%
%%%%%%%%%%%%%%%%%%%%%%%%%%%%%%%%%%%%%%%%%%%%%%%%%%%%%%%%%%%%%%%%%%%%%%%%%%%

\section{Two-derivative BMN operators and their conformal dimensions}

In this work we are concerned with operators with two covariant derivative insertions $D_\mu$ and $D_\nu$. These %%@
represent the string states $(\alpha^\mu_n)^\dagger (\alpha^\nu_{-n})^\dagger \ket{0,p^+}$ with energy
\be \label{eqn:BMN_operator_energy}
\tfrac{1}{\mu} H = 2\sqrt{1+\lambda'n^2} = 2 + \lambda' n^2 + \order{\lambda'^2} \; .
\ee
In analogy to the case of scalar field insertions we define (for $n\not=0$) 
\be \label{eqn:BMN_operator_definition}
\D_{\mu\nu,n} := \frac{J^{-1/2}}{2N_0^J} \Biggm[ \sum_{p=1}^{J-1} \tr (D_\mu Z) Z^{p-1} (D_\nu Z) Z^{J-1-p} e^{2\pi i n %%@
p/J} + \tr (D_\mu D_\nu Z) Z^{J-1} \Biggm] \; ,
\ee
where all fields are located at point $x$ and $N_0 := \sqrt{\gym^2N/8\pi^2}$. These operators carry $R$-charge $J$ and %%@
possess engineering conformal dimension
\be \label{eqn:BMN_operator_engineering_conformal_dimension}
\Delta^{(0)} = J + 2 \; .
\ee
We note that the covariant derivatives in the second term commute because of the trace. In particular the term
\be \label{eqn:derivative_of_gauge_field_unimportant}
\tr \comm{\partial_\mu A_\nu}{Z} Z^{J-1} = \tr \partial_\mu A_\nu \comm{Z}{Z^{J-1}} = 0
\ee
vanishes which will become important later since it implies that we may choose whether we want to differentiate the %%@
gauge field or rather not. As a consequence of this commutation we have the identity 
\be \label{eqn:BMN_operator_symmetry}
\D_{\mu\nu,-n} = \D_{\nu\mu,n}
\ee
and therefore we could restrict ourselves to operators with positive index $n>0$.

For the definition of the zero mode operators $(n=0)$ we take \eqref{eqn:BMN_operator_definition} with an additional %%@
normalization factor of $J^{-1}$. This is natural since $\sum e^{2\pi i n p/J}$ is of order $\order{J}$ for $n=0$ and %%@
of order $\order{1}$ otherwise. Apart from that, the definition \eqref{eqn:BMN_operator_definition} simplifies %%@
drastically in absence of the phase factor. The commutator terms cancel and the derivatives can be taken out of the %%@
trace:
\be \label{eqn:BMN_operator_zero_mode}
\begin{split}
\D_{\mu\nu,0} & := \frac{J^{-3/2}}{2N_0^J} \Biggm[ \sum_{p=1}^{J-1} \tr (D_\mu Z) Z^{p-1} (D_\nu Z) Z^{J-1-p} + \tr %%@
(D_\mu D_\nu Z) Z^{J-1} \Biggm] \\
              & = \frac{J^{-5/2}}{2N_0^J} \: \partial_\mu \partial_\nu \tr Z^J \; .
\end{split}
\ee
This means that $\D_{\mu\nu,0}$ is a descendant of the protected vacuum operator \eqref{eqn:BMN_vacuum_operator}.

The set of operators \eqref{eqn:BMN_operator_definition},\eqref{eqn:BMN_operator_zero_mode} transforms under a %%@
reducible tensor representation of $SO(4)$. A decomposition into subsets that transform under (different) irreducible %%@
representations lead to the following linear combinations ($n\in\Z$)
\begin{subequations} \label{eqn:BMN_operator_irred}
\begin{align}
\D_{(\mu\nu),n} & := \tfrac{1}{2} \bigl( \D_{\mu\nu,n} + \D_{\nu\mu,n} \bigr) - \tfrac{1}{4} \delta_{\mu\nu} %%@
\D_{\kappa\kappa,n} \; , \\ 
\D_{[\mu\nu],n} & := \tfrac{1}{2} \bigl( \D_{\mu\nu,n} - \D_{\nu\mu,n} \bigr) \; , \\
\D_{n}          & := \D_{\kappa\kappa,n} \; ,
\end{align}
\end{subequations} 
which represent the symmetric-traceless part, the anti-symmetric part and the trace, respectively.

Our aim now is to compute the conformal dimensions $\Delta_n = \Delta^{(0)} + \delta\Delta_n$ of these operators. The %%@
notation anticipates that the conformal dimensions will depend on the mode number but not on the $SO(4)$ irrep. They %%@
can be read off from the two point correlator whose generic form is
\be \label{eqn:BMN_operator_two_point_correlator}
\begin{split}
\vev{\D_{\mu\nu,n}(x)\;\bar\D_{\rho\sigma,m}(0)} & = \frac{ C_{\mu\nu,\rho\sigma}(x) \delta_{n,m} + %%@
C_{\mu\nu,\sigma\rho}(x) \delta_{n,-m} }{(x^2)^{\Delta_n}} \\
                                                 & \approx \frac{ C_{\mu\nu,\rho\sigma}(x) \delta_{n,m} + %%@
C_{\mu\nu,\sigma\rho}(x) \delta_{n,-m} }{(x^2)^{\Delta^{(0)}}} \bigl( 1 + \delta\Delta_n \ln x^{-2} \bigr) \; . 
\end{split}
\ee
Conformal invariance dictates more specifically the form of the function $C_{\mu\nu,\rho\sigma}(x)$. It is built out of %%@
Kronecker deltas $\delta_{\mu\nu}$, which connect indices referring to the same tangent space, and inversion matrices %%@
$J_{\mu\rho} \equiv \delta_{\mu\rho} - \frac{2 x_\mu x_\rho}{x^2}$, which mediate between different tangent spaces. The %%@
particular structure in the indices $n$ and $m$ means nothing but orthogonality. The term $\delta_{n,-m}$ might appear %%@
unexpected at first sight, but it is required because of the symmetry \eqref{eqn:BMN_operator_symmetry}. When %%@
restricted to $n,m\ge 0$ only the usual delta-symbol is present.

Of course it has to be checked that the correlators of the operators defined in \eqref{eqn:BMN_operator_definition} and %%@
\eqref{eqn:BMN_operator_zero_mode} indeed have the required form \eqref{eqn:BMN_operator_two_point_correlator}. In %%@
fact, we will find that for the introduced operators this is \emph{not} the case and we are compelled to redefine the %%@
non-zero mode operators \eqref{eqn:BMN_operator_definition} by adding a suitable proportion of the zero mode operators %%@
\eqref{eqn:BMN_operator_zero_mode}, cf. \eqref{eqn:operator_redefinition}.

At the technical level the one-loop anomalous dimension is determined as follows. Computing the two point correlator %%@
\eqref{eqn:BMN_operator_two_point_correlator} at tree level fixes the overall constant $C_{\mu\nu,\rho\sigma}$. %%@
Afterwards one obtains the relative factor, which essentially is the anomalous dimension, from the one-loop %%@
computation. In the latter calculation we use dimensional reduction to $d$ dimensions in order to regularize %%@
divergences and find the renormalized operators.

In what follows we will perform this calculation. We consider the BMN limit \eqref{eqn:BMN_limit} and furthermore %%@
restrict ourselves to diagrams that can be drawn on a sphere without crossing lines. As mentioned in the introduction, %%@
our result
\be
\Delta_n = J + 2 + \lambda' n^2
\ee
equals the one for operators with scalar insertions and mixed scalar-vector insertions. Due to supersymmetry arguments %%@
\cite{Beisert:2002} the matching of the conformal dimensions for operators with derivative insertions is actually %%@
expected. Nevertheless our computation provides another consistency check for the BMN proposal and we will show some %%@
specialties of our computation with regard to the scalar case.

%%%%%%%%%%%%%%%%%%%%%%%%%%%%%%%%%%%%%%%%
%%%%%%% The one-loop computation %%%%%%%
%%%%%%%%%%%%%%%%%%%%%%%%%%%%%%%%%%%%%%%%

\section{The one-loop computation} \label{sec:one_loop_computation}

%%%%%%%%%%%%%%%%%%%%%%%%%%%%%%%%%%%
%%%%%%% Preliminary remarks %%%%%%%
%%%%%%%%%%%%%%%%%%%%%%%%%%%%%%%%%%%

\subsection{Preliminary remarks}

The first thing to note is that the insertion of a covariant derivative actually produces two terms, one with partial %%@
derivative and one with the commutator of the gauge field $A$. To be more concrete let us split the operator %%@
\eqref{eqn:BMN_operator_definition} into pieces with no, one and two $A$-fields, respectively, as
\be \label{eqn:BMN_operator_decomposition}
\D_{\mu\nu,n} = d^{(0)}_{\mu\nu,n} + d^{(1)}_{\mu\nu,n} + d^{(2)}_{\mu\nu,n} \; .  
\ee
With this decomposition the two point correlator consists of 9 terms. But due to the different number of fields, it is %%@
clear that at tree level the different parts of \eqref{eqn:BMN_operator_decomposition} do not have any overlap. %%@
Moreover, since each propagator carries a factor of $\gym^2$, all three parts contribute to different orders in the %%@
coupling constant. Taking into account the normalization $N_0$, which also contains factors of $\gym$, this can be %%@
summarized by
\be
\vev{d^{(k)}_{\mu\nu,n}\;\bar d^{(l)}_{\rho\sigma,m}}_{\mbox{\scriptsize tree}} \sim (\gym^2)^k \delta_{kl} \; .
\ee
Thus, for the zeroth order we only need to determine the tree level correlator of $\D_{\mu\nu,n}$ where the covariant %%@
derivatives are replaced by partial ones. Therefore define
\be \label{eqn:classical_correlator_definition}
\vev{\D_{\mu\nu,n}\;\bar \D_{\rho\sigma,m}}_0 := \vev{d^{(0)}_{\mu\nu,n}\;\bar %%@
d^{(0)}_{\rho\sigma,m}}_{\mbox{\scriptsize tree}} \; .
\ee
The second term in \eqref{eqn:BMN_operator_decomposition} contributes to the same order as the one-loop corrections and %%@
therefore will be included below. The third term can be disregarded because its contribution to the perturbation %%@
expansion lies beyond our objective. 

In order to figure out what happens at one-loop order, we need to discuss the possible vertices. First of all there are %%@
the same interactions as in the scalar case, namely the scalar self-energy, the four scalar interaction and the gluon %%@
exchange. All of these contribute to the next-to-leading order in $\gym$ when inserted into the two point function of %%@
$d^{(0)}$ with itself. However, in our case there is also an additional contribution to this order stemming from the %%@
cross term of $d^{(0)}$ with $d^{(1)}$ which can be connected by the gluon emission vertex. Let us collect all relevant %%@
terms in the definition
\be \label{eqn:quantum_corrections_definition}
\begin{split}
\vev{\D_{\mu\nu,n}\;\bar\D_{\rho\sigma,m}}_1 := 
 &\ \vev{d^{(0)}_{\mu\nu,n}\;\bar d^{(0)}_{\rho\sigma,m}}_{\gym^2} \\ 
 &  + \vev{d^{(1)}_{\mu\nu,n}\;\bar d^{(0)}_{\rho\sigma,m}}_{\gym}
    + \vev{d^{(0)}_{\mu\nu,n}\;\bar d^{(1)}_{\rho\sigma,m}}_{\gym} \\
 &  + \vev{d^{(1)}_{\mu\nu,n}\;\bar d^{(1)}_{\rho\sigma,m}}_{\mbox{\scriptsize tree}} \; ,
\end{split}
\ee
where the subscripts $\gym^2$ and $\gym$ refer to the order of the inserted interactions.

In the next subsections we will work out \eqref{eqn:classical_correlator_definition} and %%@
\eqref{eqn:quantum_corrections_definition} in detail. For that purpose let us introduce some technicalities. For the %%@
zero mode operators the derivatives could be pulled out of the trace, cf. \eqref{eqn:BMN_operator_zero_mode}. But also %%@
the non-zero mode operators \eqref{eqn:BMN_operator_definition} can be written with the derivatives taken in front by %%@
utilizing the idea of the $q$-derivative \cite{Gursoy:2002yy}  
\be \label{eqn:BMN_operator_q_derivative}
\D_{\mu\nu,n} =  \frac{J^{-3/2}}{2N_0^J} \sum_{i,j=1}^{J} q^{j-i} \left. D^{x_i}_\mu D^{x_j}_\nu \tr Z(x_1) Z(x_2) %%@
\cdots Z(x_J) \right|_{x_1=x_2=\ldots=x}
\ee
where $q = e^{2\pi i n / J}$. Here $D^{x_i}_\mu$ acts only on $Z(x_i)$. This notation is actually nothing but the %%@
explicit version of the $q$-derivative but it is preferred in order to easier keep track which phase factor belongs to %%@
which derivative. However, it enforced us to place all fields $Z$ at different locations $x_1,\ldots,x_J$, which are %%@
taken to coincide after the derivatives are performed. This is indicated by the vertical bar. We have not yet specified %%@
where we want the gauge field that is contained in the covariant derivative to be located at. There are basically two %%@
possibilities: either at $x_i$ where it will be hit by other derivatives or at $x$ where it is protected from them. As %%@
we have argued in \eqref{eqn:derivative_of_gauge_field_unimportant} both choices yield the same result. In fact, we may %%@
more generally define
\be \label{eqn:placement_of_gauge_field}
D^{x_i}_\mu = \partial_\mu^{x_i} -i \comm{a A_\mu(x_i) + b A_\mu(x)}{\;\;\;}^{x_i}
\ee 
with arbitrary $a$ and $b$ as long as $a+b=1$. For our computation we choose $a=1$ and $b=0$ such that the gauge field %%@
\emph{is} differentiated. And we demand even more. The usual understanding is that operators only act to the right. But %%@
in order to obtain more homogeneous expressions we would like to have, in a product of covariant derivatives, all gauge %%@
fields differentiated:
\be \label{eqn:kind_of_normal_ordering}
D^{x_i}_\mu D^{x_j}_\nu \equiv \partial_\mu^{x_i} \partial_\nu^{x_j} -i \partial_\mu^{x_i} %%@
\comm{A_\nu(x_j)}{\;\;}^{x_j} -i \partial_\nu^{x_j} \comm{A_\mu(x_i)}{\;\;}^{x_i} - \comm{A_\mu(x_i)}{\;\;}^{x_i} %%@
\comm{A_\mu(x_j)}{\;\;}^{x_j} \; .
\ee
If one wishes this could be called a kind of normal ordering.

There is another important property of \eqref{eqn:BMN_operator_q_derivative}. Although the fields $Z$ have been moved %%@
to different coordinates the cyclicity of the expression is retained. Since $q^J = 1$ the coordinates can be relabeled %%@
cyclicly without changing the phase factor. Explicitly we have for some arbitrary\footnote{The only restriction is that %%@
$D_\mu^{x_i} f$ makes sense.} function $f(x_1,\ldots,x_J)$
\be \label{eqn:symmetry_in_xs}
\sum_{i,j=1}^{J} q^{j-i} \left. D^{x_i}_\mu D^{x_j}_\nu f(x_1,\ldots,x_J) \right|_{x_1=\ldots=x} = \sum_{i,j=1}^{J} %%@
q^{j-i} \left. D^{x_i}_\mu D^{x_j}_\nu f(x_2,\ldots,x_J,x_1) \right|_{x_1=\ldots=x} \; .
\ee

%%%%%%%%%%%%%%%%%%%%%%%%%%%%%%%%%%%%%%%%%%%%%%%%%%%%%
%%%%%%% Classical Correlator $\order{\gym^0}$ %%%%%%%
%%%%%%%%%%%%%%%%%%%%%%%%%%%%%%%%%%%%%%%%%%%%%%%%%%%%%

\subsection{Classical Correlator $\order{\gym^0}$}

In this section we evaluate the tree level correlator \eqref{eqn:classical_correlator_definition}. For $n,m\not=0$ it %%@
reads in terms of \eqref{eqn:BMN_operator_q_derivative}
\be \label{eqn:tree_level_start}
\vev{\D_{\mu\nu,n}\;\bar\D_{\rho\sigma,m}}_0 = \frac{J^{-3}}{4N_0^{2J}} \sum_{i,j,r,s=1}^{J} q^{j-i} p^{r-s} %%@
\partial^{x_i}_\mu \partial^{x_j}_\nu \partial^{y_r}_\rho \partial^{y_s}_\sigma \left. \vev{ \tr \prod_{k=1}^J Z(x_k) %%@
\tr \prod_{l=J}^1 \bar Z(y_l) }_{\mbox{\scriptsize tree}} \right|_{\substack{x_1=\ldots=x\\ y_1=\ldots=y}} \; ,
\ee
where $q=e^{2\pi i n/J}$ and $p=e^{2\pi i m/J}$. As before we understand the operator $\D_{\mu\nu,n}$ to be located at %%@
position $x$. The conjugated operator $\bar\D_{\rho\sigma,m}$ is implicitly understood to sit at point $y\not=x$ and %%@
their distance is denoted by $w=x-y$.

The evaluation of \eqref{eqn:tree_level_start} goes as follows. At tree level the expectation value is given by $J$ %%@
propagators that connect the two traces. The first connection can be chosen arbitrarily but the others are determined %%@
by planarity. However, using \eqref{eqn:symmetry_in_xs}, we can always relabel the coordinates in such a way that %%@
$Z(x_1)$ is connected to $\bar Z(y_1)$ and $Z(x_2)$ to $\bar Z(y_2)$ etc. This produces a factor of $J$ and another %%@
factor $N^J$ comes from the fact that the web of propagators consists of $J$ closed color lines. This leads to  
\be
\vev{\D_{\mu\nu,n}\;\bar\D_{\rho\sigma,m}}_0 = \frac{J^{-2}}{4N_0^{2J}} \left(\frac{\gym^2N}{2}\right)^J \sum_{ijrs} %%@
q^{j-i} p^{r-s} \partial^{x_i}_\mu \partial^{x_j}_\nu \partial^{y_r}_\rho \partial^{y_s}_\sigma \left. \prod_{k=1}^J %%@
I_{x_k y_k} \right|_{\substack{x_1=\ldots=x\\ y_1=\ldots=y}} \; .
\ee
The function $I_{xy}$ is the propagator defined in the appendix in \eqref{eqn:functions_propagator}. Next we carry out %%@
the derivatives and set the coordinates equal. The resulting terms that have to be summed in the following are products %%@
of $J$ propagators where some have derivatives acting onto them. As an example let us pick
\be
I_{xy}^{J-3} (\partial_\mu^x \partial_\rho^y I_{xy}) (\partial_\nu^x I_{xy}) (\partial_\sigma^y I_{xy}) = \frac{8 %%@
(4w_\mu w_\nu - \delta_{\mu\nu}w^2)w_\nu w_\sigma}{(2\pi)^{2J} (w^2)^{J+4}} \; .
\ee
This structure occurs when $i$ and $r$ are equal and $i$, $j$, and $s$ are all different from each other. Accordingly %%@
the corresponding factor is
\be
\sum_{i,j,r,s=1}^J q^{j-i} p^{r-s} \: \delta_{ir} (1-\delta_{ij}) (1-\delta_{is}) (1-\delta_{js}) = -J^2 \delta_{mn} + %%@
2J \qquad (n,m\not=0) \; .
\ee
In the cases where the indices $n$ or $m$ are zero the evaluation of the correlator using %%@
\eqref{eqn:BMN_operator_zero_mode} becomes even simpler since the derivatives can be performed after the summation. Up %%@
to terms of order $\order{J^{-1}}$ which are suppressed in the BMN limit, one finds $(n,m\not=0)$ 
\begin{subequations} \label{eqn:classical_correlator_unprimed}
\begin{align}
\vev{\D_{\mu\nu,n}\;\bar\D_{\rho\sigma,m}}_0 & = \frac{1}{(w^2)^{J+2}} \biggl[ \delta_{n,m} J_{\mu\rho} J_{\nu\sigma} + %%@
\delta_{n,-m} J_{\mu\sigma} J_{\nu\rho} \nonumber \\
                                             & \qquad\qquad\quad + \left( \delta_{\mu\nu} - \frac{2 w_\mu w_\nu}{w^2} %%@
\right) \left( \delta_{\rho\sigma} - \frac{2 w_\rho w_\sigma}{w^2} \right) \biggr] \; , \\
\vev{\D_{\mu\nu,n}\;\bar\D_{\rho\sigma,0}}_0 & = -\frac{1}{(w^2)^{J+2}} \left( \delta_{\mu\nu} - \frac{2 w_\mu %%@
w_\nu}{w^2} \right) \frac{2w_\rho w_\sigma}{w^2} \; , \\
\vev{\D_{\mu\nu,0}\;\bar\D_{\rho\sigma,m}}_0 & = -\frac{1}{(w^2)^{J+2}} \frac{2 w_\mu w_\nu}{w^2} \left( %%@
\delta_{\rho\sigma} - \frac{2 w_\rho w_\sigma}{w^2} \right) \; , \\
\vev{\D_{\mu\nu,0}\;\bar\D_{\rho\sigma,0}}_0 & = \frac{1}{(w^2)^{J+2}} \frac{4 w_\mu w_\nu w_\rho w_\sigma}{w^4} \; ,
\end{align}
\end{subequations}
where the inversion matrix $J_{\mu\rho}(w) = \delta_{\mu\rho} - 2\frac{w_\mu w_\rho}{w^2}$ has emerged. But there are %%@
also other terms that spoil the general structure \eqref{eqn:BMN_operator_two_point_correlator}. Firstly, we have found %%@
an $n$-independent piece in the correlator of the non-zero modes and, secondly, there is an overlap between non-zero %%@
and zero modes. Both troubles can be cured by the following redefinition
\be \label{eqn:operator_redefinition}
\D'_{\mu\nu,n} := \begin{cases}
  \D_{\mu\nu,n} - \D_{\mu\nu,0} + \frac{1}{2} \delta_{\mu\nu} \D_{\kappa\kappa,0} & \mbox{for $n\not=0$} \; , \\
  \D_{\mu\nu,0}                                                                   & \mbox{for $n=0$} \; .
  \end{cases}
\ee
which means in terms of the irreducible combinations \eqref{eqn:BMN_operator_irred}
\begin{align}
\D'_{(\mu\nu),n} & = \D_{(\mu\nu),n} - \D_{\mu\nu,0} \; , \\ 
\D'_{[\mu\nu],n} & = \D_{[\mu\nu],n} \; , \\
\D'_{n}          & = \D_{n} + \D_{\kappa\kappa,0} \; .
\end{align}
For these new operators we find in the BMN limit:
\begin{subequations} \label{eqn:classical_correlator_primed}
\begin{align}
\vev{\D'_{\mu\nu,n}\;\bar\D'_{\rho\sigma,m}}_0 & = \frac{\delta_{n,m} J_{\mu\rho} J_{\nu\sigma} + \delta_{n,-m} %%@
J_{\mu\sigma} J_{\nu\rho}}{(w^2)^{J+2}} \; , \label{eqn:classical_correlator_nn_primed} \\
\vev{\D'_{\mu\nu,n}\;\bar\D'_{\rho\sigma,0}}_0 & = \vev{\D'_{\mu\nu,0}\;\bar\D'_{\rho\sigma,m}}_0 = %%@
\vphantom{\frac{1}{(w^2)^{J+2}}} 0 \; , \label{eqn:classical_correlator_n0_primed} \\
\vev{\D'_{\mu\nu,0}\;\bar\D'_{\rho\sigma,0}}_0 & = \frac{4 w_\mu w_\nu w_\rho w_\sigma / w^4}{(w^2)^{J+2}} \; . %%@
\label{eqn:classical_correlator_00_primed}
\end{align}
\end{subequations}
Now the index structure of \eqref{eqn:classical_correlator_nn_primed} is exactly what conformal symmetry demands from %%@
primary operators. The particular structure of \eqref{eqn:classical_correlator_00_primed} is not especially meaningful; %%@
it is just that term of the fourth derivative of $\vev{Z \bar Z}$ that is dominant for large $J$. The only important %%@
fact here about zero mode operators is, that there is no overlap between non-zero and zero modes %%@
\eqref{eqn:classical_correlator_n0_primed}.

%%%%%%%%%%%%%%%%%%%%%%%%%%%%%%%%%%%%%%%%%%%%%%%%%%%%
%%%%%%% Quantum Corrections $\order{\gym^2}$ %%%%%%%
%%%%%%%%%%%%%%%%%%%%%%%%%%%%%%%%%%%%%%%%%%%%%%%%%%%%

\subsection{Quantum Corrections $\order{\gym^2}$}

Now we turn to the evaluation of the quantum corrections to the $\D\bar\D$-correlators %%@
\eqref{eqn:classical_correlator_primed}, which are summarized in  \eqref{eqn:quantum_corrections_definition}. To begin %%@
with we investigate only the case $n,m\not=0$; the other cases are treated in the summary on page %%@
\pageref{eqn:correlator_quantum_correction_nm_unprimed}. Explicitly we can write %%@
\eqref{eqn:quantum_corrections_definition} as
\be \label{eqn:quantum_corrections}
\vev{\D_{\mu\nu,n}\;\bar \D_{\rho\sigma,m}}_1 = \frac{J^{-3}}{4N_0^{2J}} \sum_{ijrs} q^{j-i} p^{r-s} \left. \vev{ %%@
D^{x_i}_\mu D^{x_j}_\nu D^{y_r}_\rho D^{y_s}_\sigma \tr \prod_{k=1}^J Z(x_k) \tr \prod_{l=J}^1 \bar Z(y_l) } %%@
\right|_{\substack{x_1=\ldots=x\\ y_1=\ldots=y}}
\ee
if we consider out of the 16 terms contained in $D^{x_i}_\mu D^{x_j}_\nu D^{y_r}_\rho D^{y_s}_\sigma$ only the %%@
following ones:
\begin{align} \label{eqn:quantum_corrections_relevant_terms}
& \partial^{x_i}_\mu \, \partial^{x_j}_\nu \, \partial^{y_r}_\rho \, \partial^{y_s}_\sigma           
& \longrightarrow \qquad & \vev{d^{(0)}_{\mu\nu,n}\;\bar d^{(0)}_{\rho\sigma,m}}_{\gym^2} \nonumber \\
& -i \partial^{x_j}_\nu \, \partial^{y_r}_\rho \, \partial^{y_s}_\sigma \, \comm{A_\mu}{\;\;}^{x_i}
  -i \partial^{x_i}_\mu \, \partial^{y_r}_\rho \, \partial^{y_s}_\sigma \, \comm{A_\nu}{\;\;}^{x_j}
& \longrightarrow \qquad & \vev{d^{(1)}_{\mu\nu,n}\;\bar d^{(0)}_{\rho\sigma,m}}_{\gym} \nonumber \\
& -i \partial^{x_i}_\mu \, \partial^{x_j}_\nu \, \partial^{y_s}_\sigma \, \comm{A_\rho}{\;\;}^{y_r}
  -i \partial^{x_i}_\mu \, \partial^{x_j}_\nu \, \partial^{y_r}_\rho \, \comm{A_\sigma}{\;\;}^{y_s}
& \longrightarrow \qquad & \vev{d^{(0)}_{\mu\nu,n}\;\bar d^{(1)}_{\rho\sigma,m}}_{\gym} \nonumber \\
& \begin{matrix}
  + \partial^{x_j}_\nu \, \partial^{y_s}_\sigma \, \comm{A_\mu}{\;\;}^{x_i} \, \comm{A_\rho}{\;\;}^{y_r} 
  + \partial^{x_j}_\nu \, \partial^{y_r}_\rho \, \comm{A_\mu}{\;\;}^{x_i} \, \comm{A_\sigma}{\;\;}^{y_s} \\[1mm]
  + \partial^{x_i}_\mu \, \partial^{y_s}_\sigma \, \comm{A_\nu}{\;\;}^{x_j} \, \comm{A_\rho}{\;\;}^{y_r} 
  + \partial^{x_i}_\mu \, \partial^{y_r}_\rho \, \comm{A_\nu}{\;\;}^{x_j} \, \comm{A_\sigma}{\;\;}^{y_s}
  \end{matrix}
& \left. \vphantom{\rule{1mm}{7mm}} \right\} \longrightarrow \qquad & \vev{d^{(1)}_{\mu\nu,n}\;\bar %%@
d^{(1)}_{\rho\sigma,m}}_{\mbox{\scriptsize tree}} \; . 
\end{align}
Recall the discussion concerning the placement of the gauge field $A$, cf. \eqref{eqn:placement_of_gauge_field}, and %%@
the normal ordering \eqref{eqn:kind_of_normal_ordering}. We begin with the first term in %%@
\eqref{eqn:quantum_corrections_relevant_terms}:
\be \label{eqn:quantum_corrections_first_term}
\begin{split}
\vev{d^{(0)}_{\mu\nu,n}\;\bar d^{(0)}_{\rho\sigma,m}}_{\gym^2} = & \ \frac{J^{-3}}{4N_0^{2J}} \sum_{ijrs} q^{j-i} %%@
p^{r-s} \partial^{x_i}_\mu \partial^{x_j}_\nu \partial^{y_r}_\rho \partial^{y_s}_\sigma \\
         & \left. \vev{ \tr \prod_{k=1}^J Z(x_k) \tr \prod_{l=J}^1 \bar Z(y_l) }_{\gym^2} %%@
\right|_{\substack{x_1=\ldots=x\\ y_1=\ldots=y}} \; .
\end{split}
\ee
It is similar to the tree level expression, except that we now have to evaluate the expectation value with the %%@
appropriate interactions. The required vertices are conveniently written in terms of some functions %%@
\eqref{eqn:vertex_functions} defined in the appendix. The scalar self-energy can be viewed as an interaction between %%@
two scalar fields \cite{Beisert:2002bb}
\be
\begin{split}
\raisebox{-4mm}{\includegraphics*[scale=.6]{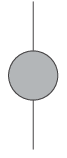}} 
  & \equiv \vev{Z^a(x_1) \bar Z^b(x_2)}_{\mbox{\scriptsize scalar self-energy}} \\ 
  & = -2 \left( \frac{\gym^2}{2} \right)^2 \bigl( N \tr T^a T^b - \tr T^a \tr T^b \bigr) \bigl( Y_{112} + Y_{122} %%@
\bigr) \; .
\end{split}
\ee
Additionally there are two interactions among four scalar fields, which are given by the 4-scalar vertex
\be \label{eqn:four_scalar_vertex}
\begin{split}
\raisebox{-4mm}{\includegraphics*[scale=.6]{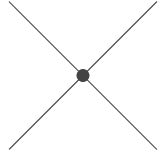}}
  & \equiv \vev{Z^a(x_1) Z^b(x_2) \bar Z^c(x_3) \bar Z^d(x_4)}_{\mbox{\scriptsize 4 scalar vertex}} \\ 
  & = - \left( \frac{\gym^2}{2} \right)^3 \bigl( \tr \comm{T^a}{T^c}\comm{T^b}{T^d} + \tr %%@
\comm{T^a}{T^d}\comm{T^b}{T^c} \bigr) X_{1234}
\end{split}
\ee
and the gluon exchange vertex
\be
\begin{split}
\raisebox{-4mm}{\includegraphics*[scale=.6]{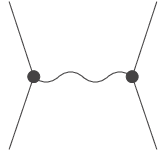}}
  & \equiv \vev{Z^a(x_1) Z^b(x_2) \bar Z^c(x_3) \bar Z^d(x_4)}_{\mbox{\scriptsize gluon exchange}} \\ 
  & = - \left( \frac{\gym^2}{2} \right)^3 \bigl( \tr \comm{T^a}{T^c}\comm{T^b}{T^d} H_{13,24} + \tr %%@
\comm{T^a}{T^d}\comm{T^b}{T^c} H_{14,23} \bigr) \; .
\end{split}
\ee
Both expressions are again taken from \cite{Beisert:2002bb}. These interactions are to be used to connect one $Z$ to %%@
one $\bar Z$ or two $Z$'s to two $\bar Z$'s in \eqref{eqn:quantum_corrections_first_term}, respectively. The others are %%@
connected by ordinary propagators. Since we are only working at planar level, the four particle interactions have to %%@
occur between adjacent scalar fields. Using again the symmetry \eqref{eqn:symmetry_in_xs} we can move the vertices to %%@
our favorite position. We choose the four modules to connect $Z(x_1)$, $Z(x_2)$, $\bar Z(y_1)$, and $\bar Z(y_2)$. And %%@
it is convenient to place half of the scalar self-energy between $Z(x_1)$ and $\bar Z(y_1)$, and the other half between %%@
$Z(x_2)$ and $\bar Z(y_2)$. Selecting only the planar contributions all three interactions can be summarized in an %%@
effective four point vertex
\be \label{eqn:v_vertex}
V_{x_1 y_1 x_2 y_2} := X_{x_1 y_1 x_2 y_2} + H_{x_1 y_1, x_2 y_2} - (Y_{x_1 x_1 y_1} + Y_{x_1 y_1 y_1}) I_{x_2 y_2} - %%@
I_{x_1 y_1} (Y_{x_2 x_2 y_2} + Y_{x_2 y_2 y_2})
\ee
or graphically
\be
\vertex{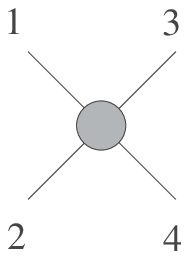}{$x_1$}{$y_1$}{$x_2$}{$y_2$}{}{}
\quad := \left( \;\; \raisebox{-4mm}{\psfrag{h}{$\frac{1}{2}$} \psfrag{p}{$+$}
                                     \includegraphics*[scale=.6]{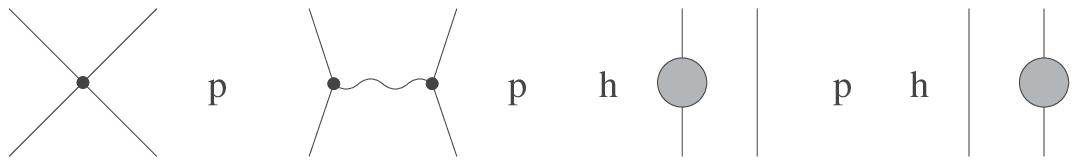}}
                \;\; \right)_{\mbox{\scriptsize planar part}} \; ,
\ee
such that \eqref{eqn:quantum_corrections_first_term} becomes
\be
\begin{split}
\vev{d^{(0)}_{\mu\nu,n}\;\bar d^{(0)}_{\rho\sigma,m}}_{\gym^2} = & \ \frac{J^{-1}}{4N_0^{2J}} \left( \frac{\gym^2N}{2} %%@
\right)^{J+1} \\
             & \times \sum_{ijrs} q^{j-i} p^{r-s} \partial^{x_i}_\mu \partial^{x_j}_\nu \partial^{y_r}_\rho %%@
\partial^{y_s}_\sigma \left[ V_{x_1 y_1 x_2 y_2} \prod_{k=3}^J I_{x_k y_k} \right]_{\substack{x_1=\ldots=x\\ %%@
y_1=\ldots=y}} \; .
\end{split}
\ee

The evaluation of this expression will simplify if we combine it with the other terms of %%@
\eqref{eqn:quantum_corrections_relevant_terms}. Therefore let us write $d_{\mu\nu,n}^{(1)}$ explicitly:
\be
\begin{split}
d_{\mu\nu,n}^{(1)} = & \frac{J^{-3/2}}{2N_0^J} \sum_{ij} q^{j-i} \partial_\nu^{x_j} \comm{(-i)A_\mu(x_i)}{\;\;}^{x_i} %%@
\left. \tr \prod_{k=1}^J Z(x_k) \right|_{\substack{x_1=\ldots=x\\ y_1=\ldots=y}} \\
                     & + (\mu \leftrightarrow \nu , q \rightarrow q^{-1}) \; .
\end{split} 
\ee
The application of the commutator to the trace produces $2J$ terms. But by appropriate relabeling of the coordinates it %%@
is seen that all of them where $A_\mu(x_i)$ is inserted behind $Z(x_i)$ (for $i=1,\ldots,J$) are actually equal and all %%@
those where $A_\mu(x_i)$ is inserted in front of $Z(x_i)$ are equal as well. This enables us to write only two terms %%@
multiplied by a factor of J
\be \label{eqn:one_A_field_commutator_applied}
\begin{split}
d_{\mu\nu,n}^{(1)} = & - \frac{J^{-1/2}}{2N_0^J} \sum_{j} q^{j} \partial_\nu^{x_j} \left. \tr Z(x_1)[ q^{-1} %%@
(-i)A_\mu(x_1) - q^{-2} (-i)A_\mu(x_2) ] \prod_{k=2}^J Z(x_k) \right|_{x_1=\ldots=x} \\
                     & + (\mu \leftrightarrow \nu , q \rightarrow q^{-1}) \; .
\end{split}
\ee
For the terms of the second line in \eqref{eqn:quantum_corrections_relevant_terms} we compute the overlap with
\be
\bar d_{\rho\sigma,m}^{(0)} = \frac{J^{-3/2}}{2N_0^J} \sum_{rs} p^{r-s} \partial_\rho^{y_r} \partial_\sigma^{y_s} %%@
\left. \tr \prod_{l=J}^1 \bar Z(y_l) \right|_{y_1=\ldots=y} \; .
\ee
The relevant interaction vertex is
\be
\begin{split}
\raisebox{-4mm}{\includegraphics*[scale=.6]{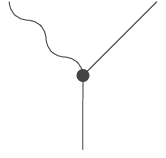}}
  & \equiv \vev{(-i)A^a_\mu(x_1) Z^b(x_2) \bar Z^c(x_3)}_{\mbox{\scriptsize gluon emission}} \\ 
  & = \left( \frac{\gym^2}{2} \right)^2 \tr T^a \comm{T^b}{T^c} \bigl( \partial^2_\mu - \partial^3_\mu \bigr) Y_{123} %%@
\; .
\end{split}
\ee
We insert it in such a way that the resulting diagrams are planar. This requires us to connect the gauge field $A$ to %%@
an adjacent $Z$, either to the left or to the right. The whole point is that these two possibilities can again be %%@
fitted nicely into an effective four point vertex
\be \label{eqn:w_vertex}
W_{\mu, x_1 y_1 x_2 y_2} := I_{x_1 y_1} \bigl( \partial_\mu^{x_2} - \partial_\mu^{y_2} \bigr) Y_{x_1 x_2 y_2} - I_{x_2 %%@
y_2} \bigl( \tfrac{1}{2} \partial_\mu^{x_1} - \partial_\mu^{y_1} \bigr) Y_{x_1 x_1 y_1} 
\ee
or graphically
\be
\vertex{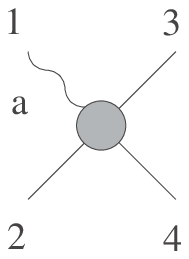}{$x_1$}{$y_1$}{$x_2$}{$y_2$}{$\mu$}{}
\quad := \left( \;\; \raisebox{-4mm}{\psfrag{p}{$+$}
                                     \includegraphics*[scale=.6]{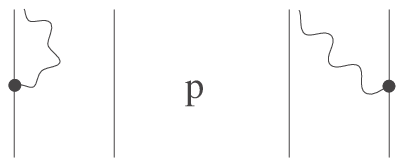}}
                \;\; \right)_{\mbox{\scriptsize planar part}} \; .
\ee
With this definition the overlap is written as
\be
\begin{split}
\vev{d^{(1)}_{\mu\nu,n}\;\bar d^{(0)}_{\rho\sigma,m}}_{\gym} = & 
\frac{J^{-1}}{4N_0^{2J}} \left( \frac{\gym^2N}{2} \right)^{J+1} \sum_{jrs} q^j p^{r-s} \partial^{x_j}_\nu %%@
\partial^{y_r}_\rho \partial^{y_s}_\sigma \\ 
        & \quad \Bigl[ \bigl( q^{-1} W_{\mu, x_1 y_1 x_2 y_2} + q^{-2} W_{\mu, x_2 y_2 x_1 y_1} \bigr) \prod_{k=3}^J %%@
I_{x_k y_k} \Bigr]_{\substack{x_1=\ldots=x\\ y_1=\ldots=y}} \\
        & + (\mu \leftrightarrow \nu , q \rightarrow q^{-1}) \; ,  
\end{split}
\ee
where a further factor of $J$ originated from singling out a particular $\bar Z$ the interaction takes place with. The %%@
third line in \eqref{eqn:quantum_corrections_relevant_terms} is given by an analogous expression.

For the terms of the last two lines in \eqref{eqn:quantum_corrections_relevant_terms} we compute the overlap of %%@
\eqref{eqn:one_A_field_commutator_applied} with itself, or more precisely with
\be
\begin{split}
\bar d_{\rho\sigma,m}^{(1)} = & - \frac{J^{-1/2}}{2N_0^J} \sum_{s} p^{-s} \partial_\sigma^{y_s} \left. \tr %%@
\prod_{l=J}^2 \bar Z(y_l)[ p^2 (-i)A_\rho(y_2) - p (-i)A_\rho(y_1) ] \bar Z(y_1) \right|_{y_1=\ldots=y} \\
                              & + (\rho \leftrightarrow \sigma , p \rightarrow p^{-1}) \; .
\end{split}
\ee
Here we obtain the order $\gym^2$ diagrams already when all fields are connected by ordinary (tree level) propagators. %%@
Nevertheless we will think of the propagators connecting $Z(x_1)$ to $\bar Z(y_1)$, $Z(x_2)$ to $\bar Z(y_2)$, and the %%@
$A$'s to each other as a four point interaction in order to unify these terms most easily with the others from above. %%@
Similar to the true interactions \eqref{eqn:v_vertex} and \eqref{eqn:w_vertex}, we therefore define the product of %%@
propagators as the following artificial four point vertex and introduce a corresponding graphical notation:
\begin{align}
R_{\mu\rho, x_1 y_1 x_2 y_2} & := \delta_{\mu\rho} I_{x_1y_1}^2 I_{x_2y_2} 
& \stackrel{\wedge}{=} &
& \vertex{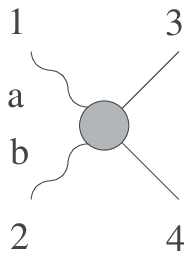}{$x_1$}{$y_1$}{$x_2$}{$y_2$}{$\mu$}{$\rho$}
\quad & := \raisebox{-4mm}{\includegraphics*[scale=.6]{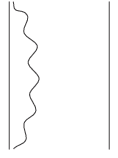}} \; ,  \label{eqn:r_vertex} \\[4mm]
U_{\mu\rho, x_1 y_1 x_2 y_2} & := -\delta_{\mu\rho} I_{x_1y_1} I_{x_1y_2} I_{x_2y_2}
& \stackrel{\wedge}{=} &
& \vertex{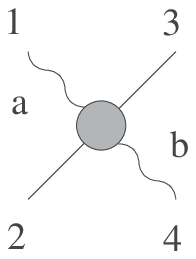}{$x_1$}{$y_1$}{$x_2$}{$y_2$}{$\mu$}{$\rho$}
\quad & := \raisebox{-4mm}{\includegraphics*[scale=.6]{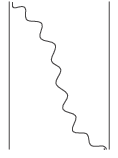}} \; .  \label{eqn:u_vertex}
\end{align}
Hence the correlator is given by
\be
\begin{split}
\vev{d^{(1)}_{\mu\nu,n}\;\bar d^{(1)}_{\rho\sigma,m}}_{\mbox{\scriptsize tree}} = & \ 
\frac{J^{-1}}{4N_0^{2J}} \left( \frac{\gym^2N}{2} \right)^{J+1} \sum_{js} q^j p^{-s} \partial^{x_j}_\nu %%@
\partial^{y_s}_\sigma \\
    & \quad \Bigl[ \bigl( q^{-1}p \, R_{\mu\rho, x_1 y_1 x_2 y_2} + q^{-1}p^2 \, U_{\mu\rho, x_1 y_1 x_2 y_2} \\
    & \quad\quad  + q^{-2}p \, U_{\mu\rho, x_2 y_2 x_1 y_1} + q^{-2}p^2 \, R_{\mu\rho, x_2 y_2 x_1 y_1} \bigr) %%@
\prod_{k=3}^J I_{x_k y_k} \Bigr]_{\substack{x_1=\ldots=x\\ y_1=\ldots=y}} \\
    & + (\mu \leftrightarrow \nu , q \rightarrow q^{-1}) + (\rho \leftrightarrow \sigma , p \rightarrow p^{-1}) \\
    & + (\mu \leftrightarrow \nu , \rho \leftrightarrow \sigma , q \rightarrow q^{-1} , p \rightarrow p^{-1}) \; .
\end{split}
\ee

\newpage % otherwise the text on the previous page is stretched too much
Now we know all relevant terms in \eqref{eqn:quantum_corrections} and list them in their pictorial version
\be \label{eqn:BMN_operator_two_point_correlator_one_loop_graphical}
\includegraphics*[scale=.64]{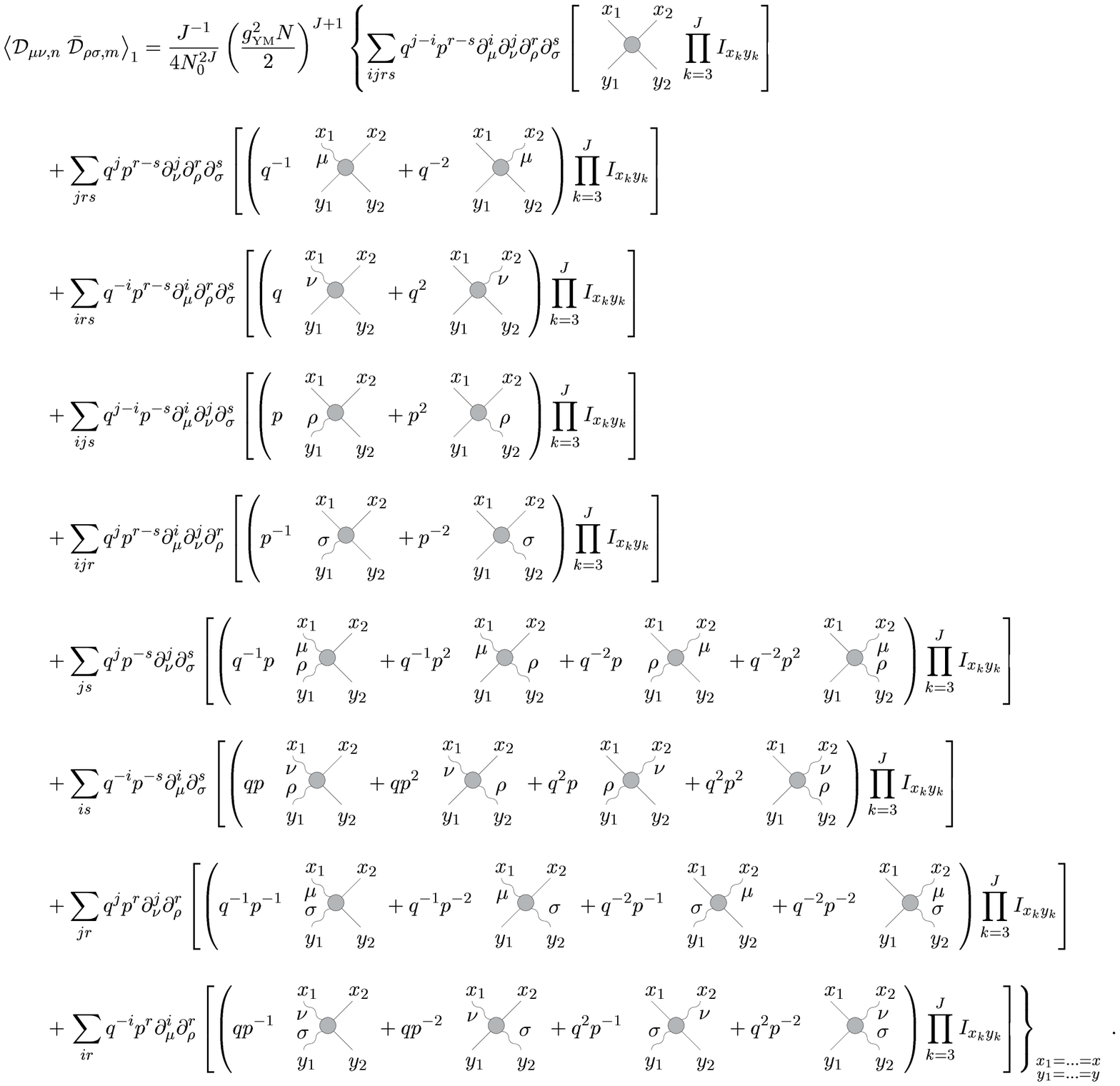}
\ee
Although this looks rather complicated it is actually pretty intuitive. We derived the last eight lines of this %%@
expression be inserting gauge fields into the chain of $Z$'s and moving them to the beginning of that chain. Afterwards %%@
the points $x_1$, $x_2$, $y_1$, and $y_2$ were connected by appropriate vertices and propagators. But effectively one %%@
may view these terms as a result of the application of some sort of operator to the legs of a four point vertex. One %%@
can further associate the same phase factors to this fictitious operator as for the derivatives that still have to be %%@
performed. This allows us to unify all lines of formula %%@
\eqref{eqn:BMN_operator_two_point_correlator_one_loop_graphical} into a single one by using the following graphical %%@
notation and definitions. 

The fields $Z(x_1)$, $Z(x_2)$, $\bar Z(y_1)$, and $\bar Z(y_2)$ are connected by some four point vertex which we will %%@
simply denote by
\be
\vertex{vertexnn}{$x_1$}{$y_1$}{$x_2$}{$y_2$}{}{} \quad .
\ee
The operator we were talking about above does not receive a new symbol but assumes that of the covariant derivative and %%@
since in the end it has the same effect we will also refer to it as covariant derivative. Hence we write %%@
\eqref{eqn:BMN_operator_two_point_correlator_one_loop_graphical} concisely as
\be \label{eqn:BMN_operator_two_point_correlator_one_loop_graphical_concise}
\begin{split}
\vev{\D_{\mu\nu,n}\;\bar \D_{\rho\sigma,m}}_1 = & \ \frac{J^{-1}}{4N_0^{2J}} \left( \frac{\gym^2N}{2} \right)^{J+1} \\
  & \times \sum_{ijrs} q^{j-i} p^{r-s} D_\mu^{x_i} D_\nu^{x_j} D_\rho^{y_r} D_\sigma^{y_s}
  \left. \left[ \quad\vertex{vertexnn}{$x_1$}{$y_1$}{$x_2$}{$y_2$}{}{}\quad \prod_{k=3}^J I_{x_k y_k} \right] %%@
\right|_{\substack{x_1=\ldots=x\\ y_1=\ldots=y}}
\end{split}
\ee
with the following meaning. The $D$'s act onto the vertex as well as onto the propagators. Applied to the latter they %%@
are just ordinary partial derivatives. But when they are applied to the vertex a sum of terms is produced consisting of %%@
partial derivatives of the effective scalar vertex \eqref{eqn:v_vertex} and of the appropriately oriented effective %%@
vector vertices \eqref{eqn:w_vertex}, \eqref{eqn:r_vertex}, and \eqref{eqn:u_vertex}. More precisely: The upper indices %%@
of the $D$'s determine the effected legs of the vertex. On these legs $D$ can either act as partial derivatives or %%@
change the legs into a gluon line. One is supposed to write all possible combinations but to discard diagrams with more %%@
than one gluon line on one side (upper two or lower two legs). For example we have 
\be
\begin{split}
D_\mu^{x_2} D_\nu^{x_2} D_\rho^{y_1} \; \vertex{vertexnn}{$x_1$}{$y_1$}{$x_2$}{$y_2$}{}{} \quad
& \equiv \vertex{vertexnn}{}{$D_\rho$}{$D_\mu D_\nu $}{}{}{} \\[3mm]
& \equiv \vertex{vertexnn}{}{$\partial_\rho$}{$\partial_\mu \partial_\nu $}{}{}{} \quad
       + \vertex{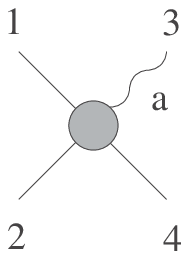}{}{$\partial_\rho$}{$\partial_\nu$}{}{$\mu$}{} \;
       + \vertex{vertexur}{}{$\partial_\rho$}{$\partial_\mu$}{}{$\nu$}{} \;
       + \vertex{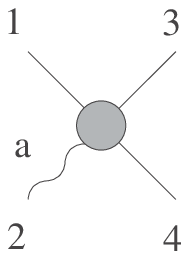}{}{}{$\partial_\mu \partial_\nu $}{}{$\rho$}{} \quad
       + \vertex{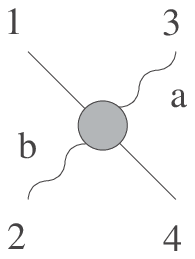}{}{}{$\partial_\nu$}{}{$\mu$}{$\rho$} \;
       + \vertex{vertexurll}{}{}{$\partial_\mu$}{}{$\nu$}{$\rho$}
        \\[3mm]
& \equiv \partial_\mu^{x_2} \partial_\nu^{x_2} \partial_\rho^{y_1} V_{x_1 y_1 x_2 y_2} \\
& \quad + \partial_\nu^{x_2} \partial_\rho^{y_1} W_{\mu, x_2 y_2 x_1 y_1} + \partial_\mu^{x_2} \partial_\rho^{y_1} %%@
W_{\nu, x_2 y_2 x_1 y_1} + \partial_\mu^{x_2} \partial_\nu^{x_2} W_{\rho, y_1 x_1 y_2 x_2} \\
& \quad + \partial_\nu^{x_2} U_{\mu\rho, x_2 y_2 x_1 y_1} + \partial_\mu^{x_2} U_{\nu\rho, x_2 y_2 x_1 y_1}\; .
\end{split}
\ee

After the derivatives have been applied all coordinates $x_k$ and $y_l$ are taken to coincide, respectively, which we %%@
depict by  
\be
\vertex{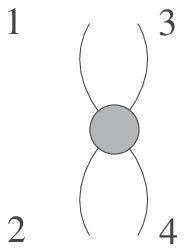}{}{\!\!$D_\rho$}{$D_\mu D_\nu $}{}{}{} \quad\quad \; .
\ee
It is helpful to perceive that vertices with joint legs have certain symmetries under reflection. Clearly, all of these %%@
vertices are symmetric under reflection about the vertical axis. Under reflection about the horizontal axis the ones %%@
with an even number of derivatives are symmetric, whereas the ones with an odd number of derivatives are %%@
anti-symmetric. The quick argument for the latter statement is, that by Lorentz covariance they can only depend on an %%@
even or odd number of $w$'s, respectively, and the reflection about the horizontal axis takes $w$ into $-w$. For %%@
instance, we have
\be
   \vertex{vertexnnjoint1}{$\!\!D_\mu$}{}{$D_\nu$}{$D_\rho$}{}{} \;\;\;
=\;\vertex{vertexnnjoint1}{$\!\!D_\nu$}{$\!\!D_\rho$}{$D_\mu$}{}{}{} \;\;\;
= -\vertex{vertexnnjoint1}{}{$\!\!D_\mu$}{$D_\rho$}{$D_\nu$}{}{} \;\;
= -\vertex{vertexnnjoint1}{$\!\!D_\rho$}{$\!\!D_\nu$}{}{$D_\mu$}{}{} \;\; \; .
\ee

In the following we wish to show the evaluation of %%@
\eqref{eqn:BMN_operator_two_point_correlator_one_loop_graphical_concise}. We will successively consider the cases where %%@
none, one, two, three and all four derivatives act on the vertex. It will turn out that only the case with exactly two %%@
derivatives leads to non-vanishing contributions. If the number of derivatives is less than two, the vertices are zero %%@
and diagrams with more than two derivatives applied to the vertex do not contribute in the BMN limit.

%%%%%%%%%%%%%%%%%%%%%%%%%%%%%%%%%%%%%%%%%%%%%
%%%%%%% Vertex without any derivative %%%%%%%
%%%%%%%%%%%%%%%%%%%%%%%%%%%%%%%%%%%%%%%%%%%%%

\subsubsection{Vertex without any derivative}

The bare vertex is given by \eqref{eqn:v_vertex} which becomes
\be
V_{xyxy} = X_{xyxy} + H_{xy,xy} - 4 Y_{xxy} I_{xy}
\ee
by joining the legs, $x_1=x_2=x$ and $y_1=y_2=y$. But this combination vanishes identically which can be seen by %%@
utilizing the following relation between $X$ and $H$ \cite{Beisert:2002bb}\footnote{This relation is written in eq. %%@
(A.7), however our function $H$ is defined to include the derivatives.}:
\be
\frac{H_{12,34}}{I_{12}I_{34}} = \frac{X_{1234}}{I_{13}I_{24}} - \frac{X_{1234}}{I_{14}I_{23}} + G_{1,34} - G_{2,34} + %%@
G_{3,12} - G_{4,12}
\ee
with
\be
G_{1,34} = \frac{Y_{134}}{I_{14}}-\frac{Y_{134}}{I_{13}} \; .
\ee
This relation simplifies in the limit $1,3\to x$ and $2,4\to y$. Taken into account that $Y_{xxy}$ and $X_{xyxy}$ %%@
contain logarithmic infinities whereas $1/I_{xx}$ is quadratically zero it is easily shown that this implies
\be \label{eqn:bare_vertex_vanishes}
\vertex{vertexnnjoint1}{}{}{}{}{}{} = V_{xyxy} = 0
\ee
and therefore
\be
\left. \vev{\D_{\mu\nu,n}\;\bar \D_{\rho\sigma,m}}_1 \right|_{\mbox{\scriptsize no derivative on vertex}} = 0 \; .
\ee

%%%%%%%%%%%%%%%%%%%%%%%%%%%%%%%%%%%%%%%%%%
%%%%%%% Vertex with one derivative %%%%%%%
%%%%%%%%%%%%%%%%%%%%%%%%%%%%%%%%%%%%%%%%%%

\subsubsection{Vertex with one derivative}

Without loss of generality let us investigate the case where the derivative sits at the upper left leg
\be \label{eqn:vertex_with_one_derivative}
\vertex{vertexnnjoint1}{$\!\!D_\mu$}{}{}{}{}{} = \vertex{vertexnnjoint1}{$\partial_\mu$}{}{}{}{}{} + %%@
\vertex{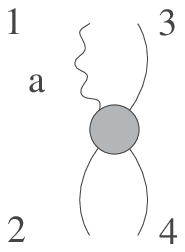}{}{}{}{}{$\mu$}{} \; .
\ee
We show that both pieces are individually zero. For the first term this is quickly seen by an enjoyable calculation %%@
invoking our graphical notation. The vertex \eqref{eqn:bare_vertex_vanishes}, where the legs are placed together, is %%@
only a function of $x$ (not $x_1$ and $x_2$ any more) and we may differentiate with respect to this coordinate. By %%@
Leibniz' rule this is related to the derivatives of the vertex before the legs were joint
\be \label{eqn:leibniz_rule_for_vertices}
\partial_\mu^x \vertex{vertexnnjoint1}{}{}{}{}{}{}
= \vertex{vertexnnjoint1}{$\partial_\mu$}{}{}{}{}{} + \vertex{vertexnnjoint1}{}{}{$\partial_\mu$}{}{}{} \; .
\ee
Now, since \eqref{eqn:bare_vertex_vanishes} vanishes for all $x$, so does its derivative, i.~e. %%@
\eqref{eqn:leibniz_rule_for_vertices} is zero. On the other hand both terms on the right hand side are equal by %%@
symmetry and therefore have to vanish as well
\be \label{eqn:one_derivative_vertex_vanishes}
\vertex{vertexnnjoint1}{$\partial_\mu$}{}{}{}{}{} = 0 \; .
\ee
(The whole argument can of course also be rigorously proven by manipulating the corresponding analytic expressions.)
The second vertex in \eqref{eqn:vertex_with_one_derivative} is immediately seen to vanish, if we set $x_1=x_2$ and %%@
$y_1=y_2$ in \eqref{eqn:w_vertex} and use $\partial_\mu^{x_2} Y_{x_1 x_2 y_2} \rightarrow \tfrac{1}{2} \partial_\mu^{x} %%@
Y_{xxy}$:
\be \label{eqn:one_gluon_vertex_vanishes}
\vertex{vertexuljoint1}{}{}{}{}{$\mu$}{} = 0 \; .
\ee
We conclude
\be
\left. \vev{\D_{\mu\nu,n}\;\bar \D_{\rho\sigma,m}}_1 \right|_{\mbox{\scriptsize one derivative on vertex}} = 0 \; .
\ee

%%%%%%%%%%%%%%%%%%%%%%%%%%%%%%%%%%%%%%%%%%%
%%%%%%% Vertex with two derivatives %%%%%%%
%%%%%%%%%%%%%%%%%%%%%%%%%%%%%%%%%%%%%%%%%%%

\subsubsection{Vertex with two derivatives}

This is the first case with non-zero vertices. As mentioned above it is the only case contributing to the anomalous %%@
dimension in the BMN limit. There are four diagrams with different placements of the two derivatives that we need to %%@
know:
\be \label{eqn:required_two_derivative_vertices}
\vertex{vertexnnjoint1}{$\!\!D_\mu$}{$\!\!D_\rho$}{}{}{}{} \quad      , \quad
\vertex{vertexnnjoint1}{$\!\!D_\mu$}{}{}{$D_\rho$}{}{}     \quad\quad , \quad
\vertex{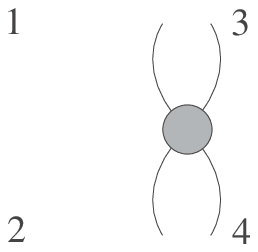}{$\!\!\!D_\mu D_\nu $}{}{}{}{}{}    \quad      , \quad
\vertex{vertexnnjoint1}{$\!\!D_\mu$}{}{$D_\nu $}{}{}{}     \quad  \;  .
\ee
Let us sketch how the computation may be done. One resolves the meaning of the vertices in terms of derivatives of the %%@
scalar and vector vertices, \eqref{eqn:v_vertex}, \eqref{eqn:w_vertex}, \eqref{eqn:r_vertex}, and \eqref{eqn:u_vertex}. %%@
Then one uses their integral representations in terms of \eqref{eqn:vertex_functions}. It is advisable to change to %%@
momentum space, where one finds a generic two loop integral of the form
\be
\int\!\frac{d^dk}{(2\pi)^d}\frac{d^dl}{(2\pi)^d}\: \frac{\ldots}{k^2 l^2 (k-p)^2 (l-p)^2 (k-l)^2} \; .
\ee
Here $d$ is the space-time dimension and $p$ the momentum of the Fourier transformation. The numerator is some function %%@
of $k$, $l$, and $p$ with two space-time indices. This kind of integral can conveniently be attacked with the aid of %%@
the mathematica package named TARCER \cite{Mertig:1998vk}. Among many other things it is capable of reducing any two %%@
loop integral into a linear combination of known basic loop integrals which have unit numerator. For example we find %%@
for the first vertex in \eqref{eqn:required_two_derivative_vertices} the following momentum space representation
\be \label{eqn:two_derivative_vertex_example_exact}
\begin{split}
\vertex{vertexnnjoint1}{$\!\!D_\mu$}{$\!\!D_\rho$}{}{}{}{} = \
 &   \frac{p_\mu p_\rho}{p^2} \
     \frac{ 3(d-4)(d-2)^2p^2 \bigl[B^{(d)}(p)\bigr]^2 - 4 (8d^3-55d^2+116d-72) J^{(d)}(p) }{3(d-4)^2(d-1)} \\
 & + \delta_{\mu\rho} \ \frac{ -3(d-4)p^2 \bigl[B^{(d)}(p)\bigr]^2 + (3d^3-13d^2+14d-16) J^{(d)}(p) }{3(d-4)^2(d-1)} \; %%@
, 
\end{split}
\ee
where
\begin{align}
B^{(d)}(p) & \equiv \int\!\frac{d^dk}{(2\pi)^d}\: \frac{1}{k^2(k-p)^2} = \frac{1}{(4\pi)^{d/2}} \frac{ %%@
\Gamma(2-\tfrac{d}{2}) \Gamma^2(\tfrac{d}{2}-1) }{\Gamma(d-2)} \frac{1}{(p^2)^{2-d/2}} \; , \\
J^{(d)}(p) & \equiv \int\!\frac{d^dk}{(2\pi)^d}\frac{d^dl}{(2\pi)^d}\: \frac{1}{k^2(l-p)^2(k-l)^2}  = %%@
\frac{1}{(4\pi)^d} \frac{ \Gamma^3(\tfrac{d}{2}-1) \Gamma(3-d) }{\Gamma(\tfrac{3d}{2}-3)} \frac{1}{(p^2)^{3-d}} \; .
\end{align}
We use the formula
\be
\int\!\frac{d^dp}{(2\pi)^d}\: \frac{e^{ipw}}{(p^2)^s} = \frac{\Gamma(\tfrac{d}{2}-s)}{4^s \pi^{d/2} \Gamma(s)} %%@
\frac{1}{(w^2)^{d/2-s}} 
\ee
for the Fourier transformation and find
\be \label{eqn:two_derivative_vertex_up_down}
\vertex{vertexnnjoint1}{$\!\!D_\mu$}{$\!\!D_\rho$}{}{}{}{} = 2 I^3(w) \, J_{\mu\rho}(w) \, L_\eps(w)
\ee
with $L_\eps(w):=-\frac{1}{\eps} + \ln w^{-2} + 1 - \gamma_E - \ln\pi$ where $\gamma_E$ is Euler's constant. Eq. %%@
\eqref{eqn:two_derivative_vertex_up_down} has been written for $d=4-2\eps$ dimensions and all terms $\order{\eps}$ have %%@
been neglected. The singular and constant (independent of $w$) terms disappear after renormalization of the operators. %%@
From \eqref{eqn:two_derivative_vertex_up_down} we can deduce the second vertex in %%@
\eqref{eqn:required_two_derivative_vertices} by a nice graphical reasoning. We differentiate %%@
\eqref{eqn:one_derivative_vertex_vanishes} and \eqref{eqn:one_gluon_vertex_vanishes} with respect to $y$
\be
0 = \partial_\rho^y \vertex{vertexnnjoint1}{$\partial_\mu$}{}{}{}{}{}
  = \vertex{vertexnnjoint1}{$\partial_\mu$}{$\partial_\rho$}{}{}{}{}
  + \vertex{vertexnnjoint1}{$\partial_\mu$}{}{}{$\partial_\rho$}{}{} \qquad , \qquad
0 = \partial_\rho^y \vertex{vertexuljoint1}{}{}{}{}{$\mu$}{}
  = \vertex{vertexuljoint1}{}{$\partial_\rho$}{}{}{$\mu$}{}
  + \vertex{vertexuljoint1}{}{}{}{$\partial_\rho$}{$\mu$}{}
\ee
and also observe
\be \label{eqn:u_r_symmetry}
  \vertex{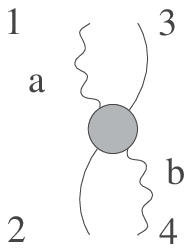}{}{}{}{}{$\mu$}{$\rho$}
  = - \delta_{\mu\rho} I_{xy}^3
  = - \vertex{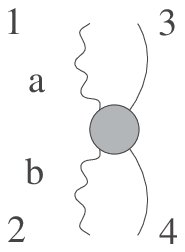}{}{}{}{}{$\mu$}{$\rho$} \; .
\ee
This implies
\be \label{eqn:two_derivative_vertex_relation}
\begin{split}
\vertex{vertexnnjoint1}{$\!\!D_\mu$}{}{}{$D_\rho$}{}{} \quad
  & =  \vertex{vertexnnjoint1}{$\partial_\mu$}{}{}{$\partial_\rho$}{}{}
      +\vertex{vertexuljoint1}{}{}{}{$\partial_\rho$}{$\mu$}{}
      +\vertex{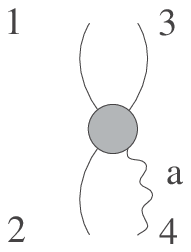}{$\partial_\mu$}{}{}{}{$\rho$}{}
      +\vertex{vertexullrjoint1}{}{}{}{}{$\mu$}{$\rho$} \\[5mm]
  & = -\vertex{vertexnnjoint1}{$\partial_\mu$}{$\partial_\rho$}{}{}{}{}
      -\vertex{vertexuljoint1}{}{$\partial_\rho$}{}{}{$\mu$}{}
      -\vertex{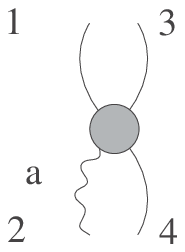}{$\partial_\mu$}{}{}{}{$\rho$}{}
      -\vertex{vertexullljoint1}{}{}{}{}{$\mu$}{$\rho$}
    = -\vertex{vertexnnjoint1}{$\!\!D_\mu$}{$\!\!D_\rho$}{}{}{}{}
    = - 2 I^3(w) \, J_{\mu\rho}(w) \, L_\eps(w) \; .
\end{split}
\ee
The third vertex in \eqref{eqn:required_two_derivative_vertices} has to be computed explicitly again. One finds an %%@
expression which is finite in four dimensions:
\be \label{eqn:two_derivative_vertex_finit}
\vertex{vertexnnjoint2}{$\!\!\!D_\mu D_\nu$}{}{}{}{} \; = I^3(w) \delta_{\mu\nu} \; .
\ee
The last vertex is the negative of the previous one which can be deduced by an argument similar to %%@
\eqref{eqn:two_derivative_vertex_relation}. In summary we list the vertices with two covariant derivatives
\begin{subequations} \label{eqn:two_derivative_vertices_summary} 
\begin{align}
     \vertex{vertexnnjoint1}{$\!\!D_\mu$}{$\!\!D_\rho$}{}{}{}{}
& = -\vertex{vertexnnjoint1}{$\!\!D_\mu$}{}{}{$D_\rho$}{}{} \; = 2 I^3(w) \, J_{\mu\rho}(w) \, L_\eps(w) \; , %%@
\label{eqn:two_derivative_vertices_summary_first_line} \\[5mm]
     \vertex{vertexnnjoint2}{$\!\!\!D_\mu D_\nu$}{}{}{}{}{}
& = -\vertex{vertexnnjoint1}{$\!\!D_\mu$}{}{$D_\nu$}{}{}{} \; = I^3(w) \, \delta_{\mu\nu} \; . %%@
\label{eqn:two_derivative_vertices_summary_second_line}
\end{align}
\end{subequations}
Now we can write down all terms in \eqref{eqn:BMN_operator_two_point_correlator_one_loop_graphical_concise} which have %%@
two derivatives acting onto the vertex. The remaining two derivatives which have to be applied to the propagators are %%@
evaluated analogously to the tree level computation. The result is some lengthy function in $J$. But since we are only %%@
interested in the BMN limit where $J\to \infty$, we may expand this function in inverse powers of $J$. One finds that %%@
the first term in this expansion is of order $\order{1/J^2}$. This is exactly what we were hoping to find because then %%@
the one-loop result possesses, compared to the tree level result, the additional factor of $\frac{N}{J^2}$, which is %%@
precisely the fixed parameter of the BMN limit \eqref{eqn:BMN_limit}, namely 
\be \label{eqn:two_derivative_vertices_result} 
\begin{split}
\left. \vev{\D_{\mu\nu,n}\;\bar \D_{\rho\sigma,m}}_1 \right|_{\substack{\mbox{\scriptsize two derivatives}\\ %%@
\mbox{\scriptsize on vertex}}} =
 & \ \frac{\gym^2 N n^2}{J^2} L_\eps(w) \frac{\delta_{n,m} J_{\mu\rho} J_{\nu\sigma} + \delta_{n,-m} J_{\mu\sigma} %%@
J_{\nu\rho}}{(w^2)^{J+2}} \\
 & - \frac{\gym^2 N n^2}{4J^2} \delta_{\mu\nu} \left( \delta_{\rho\sigma} - \frac{2 w_\rho w_\sigma}{w^2} \right) \\
 & - \frac{\gym^2 N m^2}{4J^2} \delta_{\rho\sigma} \left( \delta_{\mu\nu} - \frac{2 w_\mu w_\nu}{w^2} \right) + N %%@
\order{J^{-3}} \; .
\end{split}
\ee
The higher terms in this expansion disappear in the BMN limit, because they tend to zero faster than the single factor %%@
of $N$ in the numerator grows to infinity. Let us briefly discuss the origin of the pieces in %%@
\eqref{eqn:two_derivative_vertices_result}. The two vertices in \eqref{eqn:two_derivative_vertices_summary_first_line} %%@
or \eqref{eqn:two_derivative_vertices_summary_second_line} have the same space-time structure but receive different %%@
phase factors when inserted into \eqref{eqn:BMN_operator_two_point_correlator_one_loop_graphical_concise}, which for %%@
instance produces
\be
q^{-1} - q^{-2} = 2\pi i \frac{n}{J} + \order{J^{-2}} \; , \qquad
 p^{1} - p^{2} = -2\pi i \frac{m}{J} + \order{J^{-2}} \; .
\ee
As opposed to the scalar case where the impurities have to be connected among themselves, here it is not required that %%@
the two derivatives which are not involved in the vertex have to coincide at the same propagator. This allows terms %%@
which are off-diagonal in $n$ and $m$.

%%%%%%%%%%%%%%%%%%%%%%%%%%%%%%%%%%%%%%%%%%%%%
%%%%%%% Vertex with three derivatives %%%%%%%
%%%%%%%%%%%%%%%%%%%%%%%%%%%%%%%%%%%%%%%%%%%%%

\subsubsection{Vertex with three derivatives}

There are the following vertices
\be \label{eqn:required_three_derivative_vertices}
\vertex{vertexnnjoint1}{$\!\!D_\mu$}{$\!\!D_\rho$}{$D_\nu$}{}{}{} \quad\quad , \quad
\vertex{vertexnnjoint2}{$\!\!\!D_\mu D_\nu$}{$\!\!\!\h{D_\mu}D_\rho$}{}{}{}{} \quad , \quad
\vertex{vertexnnjoint1}{}{$\!\!D_\rho$}{$D_\mu D_\nu$}{}{}{} 
\ee
from which all vertices with three derivatives can be obtained by appropriate reflections. We can avoid evaluating the %%@
vertices explicitly if we first expand the phase factors in $1/J$. Explicitly we have
\be
\begin{split}
  & (p+p^2) \left(\vphantom{\rule{1mm}{10mm}}\right.
    \vertex{vertexnnjoint2}{$\!\!\!D_\mu D_\nu $}{$\!\!\!\h{D_\mu}D_\rho$}{}{}{}{}
  + \vertex{vertexnnjoint2}{$\!\!\!D_\mu D_\nu $}{}{}{$D_\rho$}{}{}
    \;\;\; \left.\vphantom{\rule{1mm}{10mm}}\right)
  + (qp+q^{-1}p^2) \vertex{vertexnnjoint1}{$\!\!D_\mu$}{$\!\!D_\rho$}{$D_\nu$}{}{}{}
  + (qp^2+q^{-1}p) \vertex{vertexnnjoint1}{$\!\!D_\nu$}{$\!\!D_\rho$}{$D_\mu$}{}{}{} \\[3mm]
%%%
  & = \left( 2 + 6\pi i\frac{m}{J} \right)
    \left(\vphantom{\rule{1mm}{10mm}}\right.
      \vertex{vertexnnjoint2}{$\!\!\!D_\mu D_\nu $}{$\!\!\!\h{D_\mu}D_\rho$}{}{}{}{}
  + \;\vertex{vertexnnjoint2}{$\!\!\!D_\mu D_\nu $}{}{}{$D_\rho$}{}{}\quad
  + \;\vertex{vertexnnjoint1}{$\!\!D_\mu$}{$\!\!D_\rho$}{$D_\nu$}{}{}{}\quad
  + \;\vertex{vertexnnjoint1}{$\!\!D_\mu$}{}{$D_\nu$}{$D_\rho$}{}{}
    \;\;\; \left.\vphantom{\rule{1mm}{10mm}}\right)
  + \order{J^{-2}} \; .
\end{split} 
\ee
The higher terms may be disregarded in the BMN limit taken into account the additional factor $1/J$ from the %%@
normalization and the fact that there do not emerge further powers of $J$ from the sum over the position of the fourth %%@
derivative. Thus we observe that only a particular sum of the vertices \eqref{eqn:required_three_derivative_vertices} %%@
could potentially contribute but one facilely shows that this is zero: \be \label{eqn:three_derivative_vertex_vanishes}
\begin{split}
  &   \vertex{vertexnnjoint2}{$\!\!\!D_\mu D_\nu $}{$\!\!\!\h{D_\mu}D_\rho$}{}{}{}{}
  + \;\vertex{vertexnnjoint2}{$\!\!\!D_\mu D_\nu $}{}{}{$D_\rho$}{}{} \quad
  + \;\vertex{vertexnnjoint1}{$\!\!D_\mu$}{$\!\!D_\rho$}{$D_\nu$}{}{}{} \quad
  + \;\vertex{vertexnnjoint1}{$\!\!D_\mu$}{}{$D_\nu$}{$D_\rho$}{}{} \\[3mm]
%%%
& \qquad\qquad = \partial_\rho^y \left(\vphantom{\rule{1mm}{10mm}}\right. \underbrace{
    \vertex{vertexnnjoint2}{$\!\!\!D_\mu D_\nu$ }{}{}{}{}{}
  + \vertex{vertexnnjoint1}{$\!\!D_\mu$}{}{$D_\nu$}{}{}{}
    \;\; }_{=0} \left.\vphantom{\rule{1mm}{10mm}}\right)
  + \partial_\mu^x \partial_\nu^x \left(\vphantom{\rule{1mm}{8mm}}\right. \underbrace{
    \vertex{vertexlljoint1}{}{}{}{}{$\rho$}{}
    \; }_{=0} \left.\vphantom{\rule{1mm}{8mm}}\right) \\[3mm]
%%%  
& \qquad\qquad + \partial_\mu^x \left(\vphantom{\rule{1mm}{8mm}}\right. \underbrace{
    \vertex{vertexullljoint1}{}{}{}{}{$\nu$}{$\rho$}
  + \vertex{vertexullrjoint1}{}{}{}{}{$\nu$}{$\rho$}
    \; }_{=0} \left.\vphantom{\rule{1mm}{8mm}}\right)
  + \partial_\nu^x \left(\vphantom{\rule{1mm}{8mm}}\right. \underbrace{
    \vertex{vertexullljoint1}{}{}{}{}{$\mu$}{$\rho$}
  + \vertex{vertexullrjoint1}{}{}{}{}{$\mu$}{$\rho$}
    \; }_{=0} \left.\vphantom{\rule{1mm}{8mm}}\right) \; = \: 0 \; .
\end{split} 
\ee
We have indicated which terms cancel each other by arguments given previously (cf. %%@
\eqref{eqn:two_derivative_vertices_summary_second_line}, \eqref{eqn:one_gluon_vertex_vanishes}, and %%@
\eqref{eqn:u_r_symmetry}). Hence we have 
\be
\left. \vev{\D_{\mu\nu,n}\;\bar \D_{\rho\sigma,m}}_1 \right|_{\mbox{\scriptsize three derivatives on vertex}} = 
N \order{J^{-3}} \; . 
\ee

%%%%%%%%%%%%%%%%%%%%%%%%%%%%%%%%%%%%%%%%%%%%
%%%%%%% Vertex with four derivatives %%%%%%%
%%%%%%%%%%%%%%%%%%%%%%%%%%%%%%%%%%%%%%%%%%%%

\subsubsection{Vertex with four derivatives}

In this last case, there are four essentially different vertices
\be \label{eqn:required_four_derivative_vertices}
\vertex{vertexnnjoint2}{$\!\!\!D_\mu D_\nu $}{$\!\!\!D_\rho D_\sigma $}{}{}{}{} \quad , \quad
\vertex{vertexnnjoint2}{$\!\!\!D_\mu D_\nu $}{$\!\!\!\h{D_\mu}D_\rho$}{}{$D_\sigma$}{}{} \qquad , \quad
\vertex{vertexnnjoint2}{$\!\!\!D_\mu D_\nu $}{}{}{$D_\rho D_\sigma $}{}{} \qquad\quad , \quad
\vertex{vertexnnjoint1}{$\!\!D_\mu$}{$D_\rho$}{$D_\nu$}{$D_\sigma$}{}{}
\ee
and their evaluation would be a rather tedious task. However, if we again expand the terms of %%@
\eqref{eqn:BMN_operator_two_point_correlator_one_loop_graphical_concise} with four derivatives applied to the vertex %%@
into a power series in $1/J$, we find that only the following combination of these vertices is important
\be
\begin{split}
& \quad\; \vertex{vertexnnjoint2}{$\!\!\!D_\mu D_\nu $}{$\!\!\!D_\rho D_\sigma $}{}{}{}{}\qquad
        + \vertex{vertexnnjoint2}{$\!\!\!D_\mu D_\nu $}{$\!\!\!\h{D\mu}D_\rho$}{}{$D_\sigma$}{}{}\qquad
        + \vertex{vertexnnjoint2}{$\!\!\!D_\mu D_\nu $}{$\!\!\!\h{D\mu}D_\sigma$}{}{$D_\rho$}{}{}\qquad
        + \vertex{vertexnnjoint2}{$\!\!\!D_\mu D_\nu $}{}{}{$D_\rho D_\sigma $}{}{}\qquad \\[5mm]
&       + \vertex{vertexnnjoint2}{$\!\!\!\h{D\rho}D_\mu$}{$\!\!\!D_\rho D_\sigma $}{$D_\nu$}{}{}{}\qquad
        + \quad\vertex{vertexnnjoint1}{$\!\!D_\mu$}{$\!\!D_\rho$}{$D_\nu$}{$D_\sigma$}{}{}\qquad
        + \quad\vertex{vertexnnjoint1}{$\!\!D_\mu$}{$\!\!D_\sigma$}{$D_\nu$}{$D_\rho$}{}{}\qquad
        + \quad\vertex{vertexnnjoint1}{$\!\!D_\mu$}{}{$D_\nu$}{$D_\rho D_\sigma $}{}{}\qquad \; .
\end{split}
\ee
In a couple of simple but paper consuming steps one converts this sum to
\be
\begin{split}
&   \partial_\rho^y \partial_\sigma^y \left(\vphantom{\rule{1mm}{10mm}}\right. \underbrace{
    \vertex{vertexnnjoint2}{$\!\!\!D_\mu D_\nu$ }{}{}{}{}{}
  + \vertex{vertexnnjoint1}{$\!\!D_\mu$}{}{$D_\nu$}{}{}{}
    \;\; }_{=0} \left.\vphantom{\rule{1mm}{10mm}}\right)
  + \partial_\mu^x \partial_\nu^x \partial_\sigma^y \left(\vphantom{\rule{1mm}{8mm}}\right. \underbrace{
    \vertex{vertexlljoint1}{}{}{}{}{$\rho$}{}
    \; }_{=0} \left.\vphantom{\rule{1mm}{8mm}}\right)
  + \partial_\mu^x \partial_\nu^x \partial_\rho^y \left(\vphantom{\rule{1mm}{8mm}}\right. \underbrace{
    \vertex{vertexlljoint1}{}{}{}{}{$\sigma$}{}
    \; }_{=0} \left.\vphantom{\rule{1mm}{8mm}}\right) \\[3mm]
%%%  
& + \partial_\mu^x \partial_\rho^y \left(\vphantom{\rule{1mm}{8mm}}\right. \underbrace{
    \vertex{vertexullljoint1}{}{}{}{}{$\nu$}{$\sigma$}
  + \vertex{vertexullrjoint1}{}{}{}{}{$\nu$}{$\sigma$}
    \; }_{=0} \left.\vphantom{\rule{1mm}{8mm}}\right)
  + \partial_\mu^x \partial_\sigma^y \left(\vphantom{\rule{1mm}{8mm}}\right. \underbrace{
    \vertex{vertexullljoint1}{}{}{}{}{$\nu$}{$\rho$}
  + \vertex{vertexullrjoint1}{}{}{}{}{$\nu$}{$\rho$}
    \; }_{=0} \left.\vphantom{\rule{1mm}{8mm}}\right) \\[3mm]
& + \partial_\nu^x \partial_\rho^y \left(\vphantom{\rule{1mm}{8mm}}\right. \underbrace{
    \vertex{vertexullljoint1}{}{}{}{}{$\mu$}{$\sigma$}
  + \vertex{vertexullrjoint1}{}{}{}{}{$\mu$}{$\sigma$}
    \; }_{=0} \left.\vphantom{\rule{1mm}{8mm}}\right)
  + \partial_\nu^x \partial_\sigma^y \left(\vphantom{\rule{1mm}{8mm}}\right. \underbrace{
    \vertex{vertexullljoint1}{}{}{}{}{$\mu$}{$\rho$}
  + \vertex{vertexullrjoint1}{}{}{}{}{$\mu$}{$\rho$}
    \; }_{=0} \left.\vphantom{\rule{1mm}{8mm}}\right) \; = \, 0 \; .
\end{split}
\ee
Thus we conclude that also
\be
\left. \vev{\D_{\mu\nu,n}\;\bar \D_{\rho\sigma,m}}_1 \right|_{\mbox{\scriptsize four derivatives on vertex}} = N %%@
\order{J^{-3}} \; .
\ee

%%%%%%%%%%%%%%%%%%%%%%%
%%%%%%% Summary %%%%%%%
%%%%%%%%%%%%%%%%%%%%%%%

\subsubsection{Summary}

For the non-zero mode operators $(n,m\not=0)$ we have just found
\be \label{eqn:correlator_quantum_correction_nm_unprimed}
\begin{split}
\vev{\D_{\mu\nu,n}\;\bar \D_{\rho\sigma,m}}_1 = & \ \lambda' n^2 L_\eps(w) \frac{\delta_{n,m} J_{\mu\rho} J_{\nu\sigma} %%@
+ \delta_{n,-m} J_{\mu\sigma} J_{\nu\rho}}{(w^2)^{J+2}} \\
& - \frac{\lambda' n^2}{4} \delta_{\mu\nu} \left( \delta_{\rho\sigma} - \frac{2 w_\rho w_\sigma}{w^2} \right)
  - \frac{\lambda' m^2}{4} \delta_{\rho\sigma} \left( \delta_{\mu\nu} - \frac{2 w_\mu w_\nu}{w^2} \right) \; ,
\end{split}
\ee
where all higher terms of the $1/J$ expansion have disappeared because the BMN limit was taken. We have combined the %%@
parameters according to \eqref{eqn:parameters}.
  
Let us now come to the cases where $n=0$ or $m=0$ or both. Assume first $n\not=0,m=0$ where the expression analogous to %%@
\eqref{eqn:BMN_operator_two_point_correlator_one_loop_graphical_concise} reads
\be
\vev{\D_{\mu\nu,n}\;\bar \D_{\rho\sigma,0}}_1 = \frac{J^{-2}}{4N_0^{2J}} \left( \frac{\gym^2N}{2} \right)^{J+1}
  \partial_\rho^y \partial_\sigma^y \sum_{ij} q^{j-i} D_\mu^{x_i} D_\nu^{x_j}
  \left. \left[ \quad\vertex{vertexnn}{$x_1$}{$y$}{$x_2$}{$y$}{}{}\quad \prod_{k=3}^J I_{x_k y} \right] %%@
\right|_{x_1=\ldots=x} \; .
\ee
Since we may leave the derivatives with respect to $y$ to the very end, there are only terms with none, one or two %%@
(covariant) derivatives acting onto the upper two legs of the vertex. As a consequence the only contribution stems from %%@
\eqref{eqn:two_derivative_vertices_summary_second_line} and produces
\be
\vev{\D_{\mu\nu,n}\;\bar \D_{\rho\sigma,0}}_1 = \frac{\lambda' n^2}{4} \delta_{\mu\nu} \frac{2 w_\rho w_\sigma}{w^2} \; %%@
.
\ee
If however $n=0$ and $m\not=0$, the result is 
\be
\vev{\D_{\mu\nu,0}\;\bar \D_{\rho\sigma,m}}_1 = \frac{\lambda' m^2}{4} \delta_{\rho\sigma} \frac{2 w_\mu w_\nu}{w^2} \; %%@
.
\ee
For $n=m=0$ there is no quantum correction at all since in this case we have a correlator of two (descendants of) %%@
protected operators. But this is also easy to comprehend from the computation since the vertex without derivatives has %%@
been shown to vanish:
\be \label{eqn:correlator_quantum_correction_00_unprimed}
\vev{\D_{\mu\nu,0}\;\bar \D_{\rho\sigma,0}}_1 = 0 \; .
\ee \\ % new paragraph

Finally it remains to apply the transformation \eqref{eqn:operator_redefinition} in order to find the correlators of %%@
the redefined operators $(n,m\not=0)$:
\begin{subequations} \label{eqn:correlator_quantum_correction_primed}
\begin{align}
\vev{\D'_{\mu\nu,n}\;\bar\D'_{\rho\sigma,m}}_1 & = \lambda' n^2 L_\eps(w) \frac{\delta_{n,m} J_{\mu\rho} J_{\nu\sigma} %%@
+ \delta_{n,-m} J_{\mu\sigma} J_{\nu\rho}}{(w^2)^{J+2}} \; , \\
\vev{\D'_{\mu\nu,n}\;\bar\D'_{\rho\sigma,0}}_1 & = \frac{\lambda' n^2}{4} \delta_{\mu\nu} \frac{2 w_\rho w_\sigma}{w^2} %%@
\; , \\
\vev{\D'_{\mu\nu,0}\;\bar\D'_{\rho\sigma,m}}_1 & = \frac{\lambda' m^2}{4} \delta_{\rho\sigma} \frac{2 w_\mu w_\nu}{w^2} %%@
\; , \\
\vev{\D'_{\mu\nu,0}\;\bar\D'_{\rho\sigma,0}}_1 & = 0 \; .
\end{align}
\end{subequations}
In this way we again got rid of the unwanted terms in \eqref{eqn:correlator_quantum_correction_nm_unprimed}. However, %%@
we could not abolish the overlap between non-zero and zero mode operators, which is basically due to the fact that %%@
\eqref{eqn:correlator_quantum_correction_00_unprimed} vanishes and cannot aid in the transformation %%@
\eqref{eqn:operator_redefinition}. It is therefore necessary to perform another redefinition. This time it will involve %%@
the coupling constant $\lambda'$, since the plan is to make use of the tree level correlator %%@
\eqref{eqn:classical_correlator_00_primed} in order to remove the overlap at one-loop level. The suitable redefinition %%@
is ($n\not=0$)
\be \label{eqn:second_operator_redefinition}
\D''_{\mu\nu,n} := \D'_{\mu\nu,n} - \frac{1}{8} \lambda' n^2 \delta_{\mu\nu} \D'_{\kappa\kappa,0}
                = \D_{\mu\nu,n} - \D_{\mu\nu,0} + \frac{1}{2} \delta_{\mu\nu} \left( 1 - \frac{\lambda' n^2}{4} \right) %%@
\D_{\kappa\kappa,0}
\ee
and leads to
\begin{subequations} \label{eqn:correlator_quantum_correction_doubleprimed}
\begin{align}
\vev{\D''_{\mu\nu,n}\;\bar\D''_{\rho\sigma,m}}_1 & = \lambda' n^2 L_\eps(w) \frac{\delta_{n,m} J_{\mu\rho} %%@
J_{\nu\sigma} + \delta_{n,-m} J_{\mu\sigma} J_{\nu\rho}}{(w^2)^{J+2}} \; , \\
\vev{\D''_{\mu\nu,n}\;\bar\D''_{\rho\sigma,0}}_1 & =
\vev{\D''_{\mu\nu,0}\;\bar\D''_{\rho\sigma,m}}_1 =
\vev{\D''_{\mu\nu,0}\;\bar\D''_{\rho\sigma,0}}_1 = 0 \; .
\end{align}
\end{subequations}
The tree level correlators are not affected by this redefinition.

%%%%%%%%%%%%%%%%%%%%%%%%%%%%%%%%%%%
%%%%%%% Anomalous dimension %%%%%%%
%%%%%%%%%%%%%%%%%%%%%%%%%%%%%%%%%%%

\subsection{Anomalous dimension}

We add the quantum corrections \eqref{eqn:correlator_quantum_correction_doubleprimed} to the classical part of the %%@
correlators \eqref{eqn:classical_correlator_primed}:
\begin{align}
\vev{\D''_{\mu\nu,n}\;\bar\D''_{\rho\sigma,m}} & = \frac{\delta_{n,m} J_{\mu\rho} J_{\nu\sigma} + \delta_{n,-m} %%@
J_{\mu\sigma} J_{\nu\rho}}{(w^2)^{J+2}} \left( 1 + \lambda' n^2 L_\eps(w) \right) \; , \\
\vev{\D''_{\mu\nu,0}\;\bar\D''_{\rho\sigma,0}} & = \frac{4 w_\mu w_\nu w_\rho w_\sigma / w^4}{(w^2)^{J+2}}
\end{align}
and without overlap between non-zero and zero modes. Recall that this result is only valid in the BMN limit and that it %%@
represents only the planar part of the correlator. In order to read off the anomalous dimension we actually have to %%@
renormalize the non-zero mode operators by
\be \label{eqn:BMN_operator_renormalized}
\D^{\rm ren}_{\mu\nu,n} := \D''_{\mu\nu,n}\left( 1 + \lambda' f(\eps) \right)  \qquad\mbox{with}\qquad  f(\eps) = %%@
\frac{n^2}{2} \left( \frac{1}{\eps} - 1 + \gamma_E + \ln\pi \right) \; .
\ee
This only amounts to the replacement $L_\eps(w) \to \ln w^{-2}$. Thus, by comparison with %%@
\eqref{eqn:BMN_operator_two_point_correlator}, we find the anomalous dimension
\be
\delta\Delta_n = \lambda' n^2
\ee
for all $n\in\Z$. Note that this result is independent of the particular $\grSO(4)$ irrep. Together with %%@
\eqref{eqn:BMN_operator_engineering_conformal_dimension} we obtain
\be
\Delta_n - J = \Delta^{(0)} - J + \delta\Delta_n = 2 + \lambda' n^2  
\ee
in accordance with \eqref{eqn:BMN_operator_energy}.

%%%%%%%%%%%%%%%%%%%%%%%%%%
%%%%%%% Conclusion %%%%%%%
%%%%%%%%%%%%%%%%%%%%%%%%%%

\section{Conclusion} \label{sec:conclusion}

Our aim was to find BMN operators that represent the plane-wave string states
\be \label{eqn:conclusion_string_state}
(\alpha^\mu_n)^\dagger (\alpha^\nu_{-n})^\dagger \ket{0,p^+}
\ee
where the indices $\mu,\nu$ belong to that $\grSO(4)$ subgroup of the symmetries of the plane-wave background that used %%@
to be symmetries of the $AdS_5$ space before the BMN limit was taken. These states are generically modeled by operators %%@
with covariant derivative insertions \cite{Berenstein:2002jq}. In analogy to the scalar case we defined in %%@
\eqref{eqn:BMN_operator_definition} and \eqref{eqn:BMN_operator_zero_mode} ($n\not=0$):
\begin{align}
\D_{\mu\nu,n} & := \frac{J^{-1/2}}{2N_0^J} \Biggm[ \sum_{p=1}^{J-1} \tr (D_\mu Z) Z^{p-1} (D_\nu Z) Z^{J-1-p} e^{2\pi i %%@
n p/J} + \tr (D_\mu D_\nu Z) Z^{J-1} \Biggm] \; , \label{eqn:conclusion_op_n} \\
\D_{\mu\nu,0} & := \frac{J^{-5/2}}{2N_0^J} \: \partial_\mu \partial_\nu \tr Z^J \; . \label{eqn:conclusion_op_0}
\end{align}
While the zero mode operators are explicitly descendants of a primary operator, it turned out that the non-zero mode %%@
operators are \emph{not} primary. We showed that they become primary operators by a slight modification %%@
\eqref{eqn:operator_redefinition}:
\be \label{eqn:conclusion_op_primed}
\begin{split}
\D'_{\mu\nu,n} := \frac{J^{-1/2}}{2N_0^J} \Biggm[ & \sum_{p=1}^{J-1} \tr (D_\mu Z) Z^{p-1} (D_\nu Z) Z^{J-1-p} e^{2\pi %%@
i n p/J} + \tr (D_\mu D_\nu Z) Z^{J-1} \\
                                                  & + \left( \tfrac{1}{2} \delta_{\mu\nu} \square - \partial_\mu %%@
\partial_\nu \right) \frac{1}{J^2} \tr Z^J \Biggm] \; .
\end{split}
\ee
These operators, \eqref{eqn:conclusion_op_primed} and \eqref{eqn:conclusion_op_0}, then had proper two point %%@
correlators from a conformal field theory point of view. However, at one-loop level the non-zero mode operators were %%@
not orthogonal to the zero mode operators. This required a second redefinition %%@
\eqref{eqn:second_operator_redefinition}, which this time had to depend on the coupling constant $\lambda'$. This might %%@
seem strange at first sight since this kind of redefinition is not required in the case of scalar impurities, where %%@
zero mode and non-zero mode operators do not overlap. This is because of the fact that the zero mode operator with %%@
scalar impurities is a primary operator whereas the one with derivative impurities is a descendent operator. But apart %%@
from that one should keep in mind that there is generally -- also in the scalar case -- a $\lambda'$ dependent %%@
redefinition, namely when the operators are renormalized, cf. \eqref{eqn:BMN_operator_renormalized}. Including all %%@
redefinitions we have found that \eqref{eqn:conclusion_string_state} can be represented by
\begin{subequations} \label{eqn:conclusion_op_renormalized}
\begin{align}
\D^{\rm ren}_{\mu\nu,n} & := \left(1+\lambda'f(\eps)\right) \frac{J^{-1/2}}{2N_0^J} \Biggm[ \sum_{p=1}^{J-1} \tr (D_\mu %%@
Z) Z^{p-1} (D_\nu Z) Z^{J-1-p} e^{2\pi i n p/J} + \tr (D_\mu D_\nu Z) Z^{J-1} \nonumber \\
                        & \qquad\qquad\qquad\qquad\qquad\quad + \left( \tfrac{1}{2} \delta_{\mu\nu} \square - %%@
\partial_\mu \partial_\nu \right) \frac{1}{J^2} \tr Z^J - \frac{\lambda'n^2}{8} \delta_{\mu\nu} \square \frac{1}{J^2} %%@
\tr Z^J \Biggm] \; , \\
\D_{\mu\nu,0}           & := \frac{J^{-5/2}}{2N_0^J} \: \partial_\mu \partial_\nu \tr Z^J \; .
\end{align}
\end{subequations}
Their anomalous dimensions give exactly the masses of \eqref{eqn:conclusion_string_state} at order $\order{\lambda'}$.

Actually this kind of operators had been studied before. In \cite{Beisert:2002} N.~Beisert derived operators %%@
representing the state \eqref{eqn:conclusion_string_state} by arguments using superconformal symmetry. They look %%@
different for different $\grSO(4)$ irreps. The exponential phase factors are replaced by sine and cosine functions. %%@
These operators are well-defined also outside the BMN limit for finite $J$. However for the limit \eqref{eqn:BMN_limit} %%@
where the BMN correspondence is supposed to hold, we have shown that \eqref{eqn:conclusion_op_renormalized} represents %%@
\eqref{eqn:conclusion_string_state}, at least up to one-loop order and in the planar limit.

A practical reason prefers \eqref{eqn:conclusion_op_renormalized} to the finite J BMN operators of \cite{Beisert:2002}. %%@
If one wants to actually compute the anomalous dimensions from diagrams it proves advantageous to utilize the %%@
$q$-derivative of \cite{Gursoy:2002yy}. But then we need a phase factor $q$ with $q^J=1$ in order to retain the %%@
cyclicity of the trace and hence we are led to exponentials. 

As we have shown at length in section \ref{sec:one_loop_computation} the form of the operators allowed us to write the %%@
two point correlation function in a way convenient for the one-loop calculation, cf. %%@
\eqref{eqn:BMN_operator_two_point_correlator_one_loop_graphical_concise}. The details are rather intricate but %%@
effectively it amounts to a correlator of $\tr Z^J$ and $\tr \bar Z^J$ with one generic vertex and four derivative %%@
operations acting onto the whole object. Working with this expression in the subsequent computations consisted in %%@
simple graphical manipulations following merely from Leibniz' rule and reflection symmetries. The results were the %%@
following. If none or one derivative act onto the vertex, the vertex vanishes. If three or all four derivatives act %%@
onto the vertex, the diagrams are negligible in the BMN limit. Only the case where two derivatives act onto the vertex %%@
contributes to the anomalous dimension.

Regrettably, eq. \eqref{eqn:BMN_operator_two_point_correlator_one_loop_graphical_concise} could only be found so easily %%@
in the planar limit. Though one could find effective vertices in general, all the nice and valuable relations between %%@
them only hold in the planar case. Moreover, in our planar computation we could move all different interactions on top %%@
of each other (always occurring between $Z(x_1)$, $Z(x_2)$, $\bar Z(y_1)$ and $\bar Z(y_1)$) and see them frequently %%@
cancel each other. This will not be possible in general and one has a large number of effective vertices. And as %%@
opposed to the case of scalar impurities, this cancellation is believed not to take place before the summation over the %%@
impurity insertion points. This is basically due to an essential difference between derivative and scalar case, namely %%@
the fact that derivative impurities have non-vanishing overlap with the ``background'' field $Z$, whereas scalar %%@
impurities do not:
\be
\vev{D_\mu Z(x) \; \bar Z(y)} \not= 0 \qquad;\qquad \vev{\phi_i(x) \; \bar Z(y)} = 0 \; .
\ee

It is left for the future to find the subset of all vertices that are in the end relevant for computations involving %%@
two derivative BMN operators. This will hopefully be the analogue of the effective scalar vertex such that the symmetry %%@
between the two cases would be more apparent.

%%%%%%%%%%%%%%%%%%%%%%%%%%%%%%%
%%%%%%% Acknowledgments %%%%%%%
%%%%%%%%%%%%%%%%%%%%%%%%%%%%%%%

\acknowledgments

My special thanks go to Jan Plefka and Niklas Beisert for introducing me to the topic, for valuable discussions and %%@
useful hints. For further helpful discussions I would like to thank Ari Pankiewicz, Matthias Staudacher, Charlotte %%@
Kristjansen and Markus P{\"o}ssel.

%%%%%%%%%%%%%%%%%%%%%%%%
%%%%%%% Appendix %%%%%%%
%%%%%%%%%%%%%%%%%%%%%%%%
\appendix

%%%%%%%%%%%%%%%%%%%%%%%%%%%%%%%%%%%%%%%%%%%
%%%%%%% Appendix: Yang-Mills Theory %%%%%%%
%%%%%%%%%%%%%%%%%%%%%%%%%%%%%%%%%%%%%%%%%%%

\section{Appendix: Yang-Mills theory} \label{sec:appendix_yang_mills}

We use the following Euclidean action of $\mathcal{N}=4$ supersymmetric Yang-Mills theory in $d=4$ dimensions:  
\be
\begin{split}
S = \frac{2}{\gym^2} \int\!d^4x\: \tr \Biggl( & \frac{1}{4} F_{\mu\nu}F_{\mu\nu} + \frac{1}{2}D_\mu \phi_I D_\mu \phi_I %%@
- \frac{1}{4} \comm{\phi_I}{\phi_J} \comm{\phi_I}{\phi_J} \\ 
                                               & + \frac{1}{2} \bar\psi \Gamma_{\!\mu} D_\mu \psi - \frac{i}{2} %%@
\bar\psi \Gamma_{\!I} \comm{\phi_I}{\psi} \Biggr) \; ,
\end{split}
\ee
which we fix in Feynman gauge. The quantum fields are the gauge field $A_\mu^a$ ($\mu = 1,\ldots,4$), six scalars %%@
$\phi_I^a$ ($I=1,\ldots,6$) and four Majorana spinors $\psi^a$. The field strength is given by $F_{\mu\nu} = %%@
\partial_\mu A_\nu - \partial_\nu A_\mu -i \comm{A_\mu}{A_\nu}$ and the covariant derivative by $D_\mu = \partial_\mu - %%@
i \comm{A_\mu}{\;\;}$. All fields transform in the adjoint representation of the gauge group, which in our case is %%@
$\grU(N)$. The index $a$ labels the generators $T^a$ of $\grU(N)$ and assumes the values $0,\ldots,N^2-1$. The %%@
generators obey the Lie algebra
\be
\comm{T^a}{T^b} = i f^{abc} T^c
\ee
and are normalized to
\be
\tr T^a T^b = \delta^{ab} \; , \qquad 
\sum_{a=0}^{N^2-1} (T^a)_{\alpha\beta} (T^a)_{\gamma\delta} = \delta_{\alpha\delta} \delta_{\beta\gamma} \; . 
\ee

In the BMN correspondence one singles out a $\grU(1) \cong \grSO(2)$ subgroup of the $\grSU(4) \cong \grSO(6)$ %%@
R-symmetry group. Label the scalar fields such that this $\grU(1)$ subgroup rotates $\phi_5$ and $\phi_6$ into each %%@
other. Then $Z:=\frac{1}{\sqrt{2}}(\phi_5+i\phi_6)$ carries unit charge with respect to this $\grU(1)$. The complex %%@
conjugated field $\bar Z$ possesses charge $-1$, whereas the remaining 4 scalar fields are neutral. \\

The propagators and vertices are conveniently written in terms of the following functions which were defined in %%@
\cite{Beisert:2002bb}
\begin{subequations} \label{eqn:vertex_functions} 
\begin{align} 
I_{12}    & := \int\!\frac{d^dp}{(2\pi)^d} \frac{e^{ip(x_1-x_2)}}{p^2} = \frac{1}{4\pi^{d/2}} %%@
\frac{\Gamma(\tfrac{d}{2}-1)}{[(x_1-x_2)^2]^{d/2-1}} \;\xrightarrow{\;d \to 4\;}\; \frac{1}{4\pi^2} %%@
\frac{1}{(x_1-x_2)^2} \; , \label{eqn:functions_propagator} \\
Y_{123}   & := \int\!d^du\: I_{1u} I_{2u} I_{3u} \; , \\
X_{1234}  & := \int\!d^du\: I_{1u} I_{2u} I_{3u} I_{4u} \; , \\
H_{12,34} & := (\partial^1 - \partial^2) \cdot (\partial^3 - \partial^4) \int\!d^du\,d^dv\: I_{1u}  I_{2u} I_{uv} %%@
I_{v3} I_{v4} \; .
\end{align}
\end{subequations}
The arguments have been written as indices with the meaning $1 \equiv x_1$ etc. The propagators for scalar fields and %%@
the gauge field are
\begin{align}
\vev{Z^a(x)    \bar Z^b(y)} & = \frac{\gym^2}{2} \tr T^a T^b \: I_{xy} \; , \\
\vev{A^a_\mu(x) A^b_\nu(y)} & = \frac{\gym^2}{2} \delta_{\mu\nu} \tr T^a T^b \: I_{xy} \; ,
\end{align}
respectively. The required vertices are given in the main text.

%%%%%%%%%%%%%%%%%%%%%%%%%%
%%%%%%% References %%%%%%%
%%%%%%%%%%%%%%%%%%%%%%%%%%

\end{document}